\documentclass[5p,sort&compress,10pt,fleqn]{elsarticle}
\usepackage{amsmath,amssymb}
\usepackage{graphicx}
\usepackage{xcolor}
\usepackage{xfrac}
\usepackage[colorlinks=True]{hyperref}
\usepackage{cleveref}
\usepackage{booktabs}
\usepackage{multicol}

\journal{Nuclear Instruments and Methods in Physics Research}

\begin{document}

\begin{frontmatter}


\title{Isotopic gamma lines for identification of shielding materials}
\author[1]{Oleg Korobkin}
\ead{korobkin@lanl.gov}
\author[1]{Marc L.~Klasky}
\author[2]{Ajeeta Khatiwada}
\author[1]{Michael McCann}
\affiliation[1]{
  organization={Applied Maths and Plasma Physics (T-5), Los Alamos National Laboratory}, 
  city={Los Alamos}, 
  postcode={87545},
  state={NM},
  country={USA}}
\affiliation[2]{
  organization={Materials and Physical Data (XCP-5), Los Alamos National Laboratory}, 
  city={Los Alamos}, 
  postcode={87545},
  state={NM},
  country={USA}}

\begin{abstract}

Identifying the constituting materials of concealed objects is crucial in a wide range of sectors, such as medical imaging, geophysics, nonproliferation, national security investigations, and so on.
Existing methods face limitations, particularly when multiple materials are involved or when there are challenges posed by scattered radiation and large areal mass.
Here we introduce a novel brute-force statistical approach for material identification using high spectral resolution detectors, such as HPGe.
The method relies upon updated semianalytic formulae for computing uncollided flux from source of gamma radiation, shielded by a sequence of nested spherical or cylindrical materials. 
These semianalytical formulae make possible rapid flux estimation for material characterization via combinatorial search through all possible combinations of materials, using a high-resolution HPGe counting detector.
An important prerequisite for the method is that the geometry of the objects is known (for example, from X-ray radiography). 
We demonstrate the viability of this material characterization technique in several use cases with both simulated and experimental data.

\end{abstract}


\end{frontmatter}

\section{Introduction}
\label{sec:intro}

The use of penetrating radiation to perform material identification is valuable in a variety of medical, baggage screening, geophysics, industrial, nonproliferation and nuclear security applications   ~\cite{alvarez1976,kalender1986,vetter1986,yingz2006,fraser1986,paziresh2016,cann1982}. While significant advances in material identification have been achieved using dual-energy computed tomography in both the medical and baggage scanning applications, the industrial and nuclear security domain settings offer additional challenges that demand new technologies and algorithms~\cite{alvarez1976,runkle2009}.
Among the major challenges in addressing material identification in the nuclear security and industrial settings is the necessity to examine thick objects, i.e., those with large areal mass.
For such objects, even for a low effective nuclear charge ($Z_{\rm eff}$) material composition, X-rays that are typically used in medical and baggage-screening applications have energy spectra too low (70--100 and 135-150 kVp) to penetrate the object.
Furthermore, the presence of significant scattered radiation that generally occurs in the security and industrial settings, due to the larger areal mass, brings about additional challenges that we seek to address in this work.

Researchers have attempted to address these additional challenges by employing more penetrating radiations, i.e. those with MeV scale~\cite{ogorodnikov2002,naydenov2004}. 
However, imaging in this energy regime suffers from minimal contribution from the photoelectric effect, whose $Z^4$ scaling is largely responsible for the ability to perform material identification. 
Moreover, in this energy regime, attenuation is nearly constant across both the Compton window and the onset of the pair production. Attempts to perform material identification exploiting the $Z^2$-dependence at high energy, i.e. 10 MeV, is possible, but these sources are generally not available in the industrial and security arenas, and the overlap in spectra using these poly-energetic spectra is significant.
Finally, in both industrial and nuclear security applications the X-ray may traverse multiple materials, further limiting the application of traditional dual energy CT algorithms, only providing an average $Z_{\rm eff}$ and not isolation of the material composition of the constituents~\cite{tobias2019,runkle2009}.

Recent advances in detector technology, i.e. counting detectors, have opened up a potentially promising route to enhance the ability to perform material identification in nuclear security and industrial settings~\cite{beldjoudi2012,wux2017}. It has also been recognized that monoenergetic sources offer greater enhancement in performance relative to polychromatic sources~\cite{mendoncca2013, xuey2019, mendoncca2010, long2014, kelcz1979}.  However, these high-energy monoenergetic sources are not readily available~\cite{runkle2009}. Finally, if prior information regarding the interfaces of materials, via segmentation of a 3D object, can be obtained, a discrete material identification problem may be formulated~\cite{herman2007}.  Indeed, this prior has recently been used to enable material identification via a single integrating flat-panel detector~\cite{mccann2023}.

The objective of this work is to use spectroscopic information obtained from counting detectors in conjunction with multiple emission lines and prior geometric information that may be obtained from a traditional single image integrating radiographic detector to enable material identification in a nuclear security application. 
This specific application is one in which each segmented region of a radiographic image contains only a single material with a known density and associated attenuation coefficient. An illustration of one such object is depicted in Figure~\ref{fig:problem-formulation}. We demonstrate our proposed algorithm using both simulations as well as experimental data on several spherically symmetric geometries with multiple layers of materials. 

Our specific contributions are \begin{enumerate}
\item the formulation of a novel and important material identification problem;
\item development of semi-analytic expressions for un-collided flux for numerous configurations;
\item a novel algorithm to enable rapid multi-material identification;
and
\item  validation of the algorithm on synthetic as well as experimental data.
\end{enumerate}

This article is organized as follows:  Section \ref{sec:problem} outlines our material identification problem using monoenergetic line emissions from radioactive sources in conjunction with a counting detector and segmented objects. Section \ref{sec:statmodel} presents the statistical method used for finding the optimal material configuration. In Section \ref{sec:analytic}, we outline the mathematical formalism for calculating un-collided flux to enable our material identification algorithm. Section \ref{sec:results} contains applications and case studies for various configurations. Conclusions are presented in Section \ref{sec:conclusion}.

\section{Problem Formulation}
\label{sec:problem}

\begin{figure*}[!htbp]
    \centering
    \begin{tabular}{ccc}
    \includegraphics[width=0.34\textwidth]{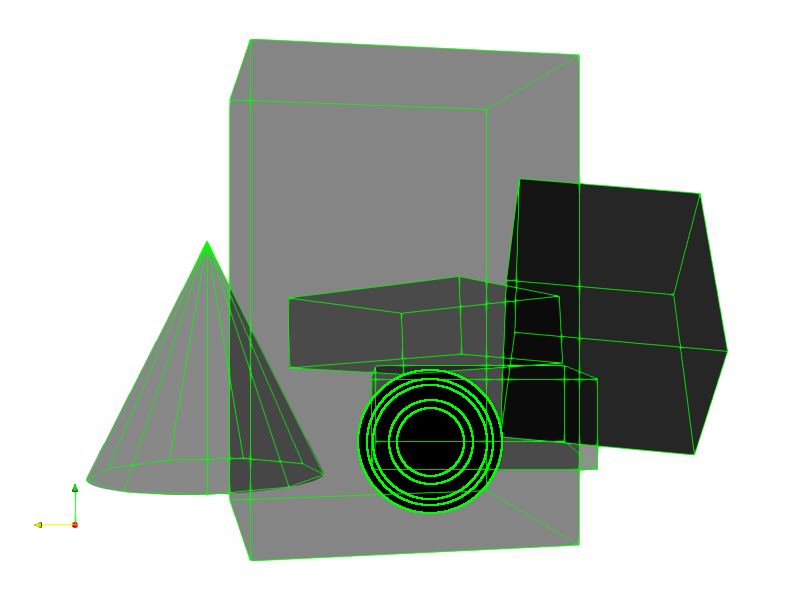} &
    \includegraphics[width=0.20\textwidth]{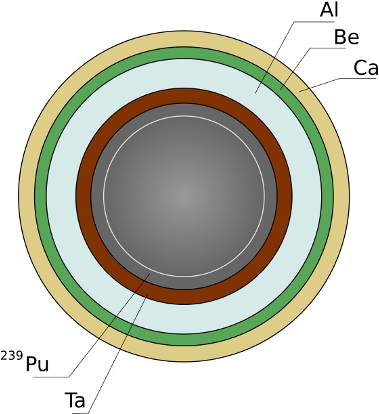} &  
    \includegraphics[width=0.35\textwidth]{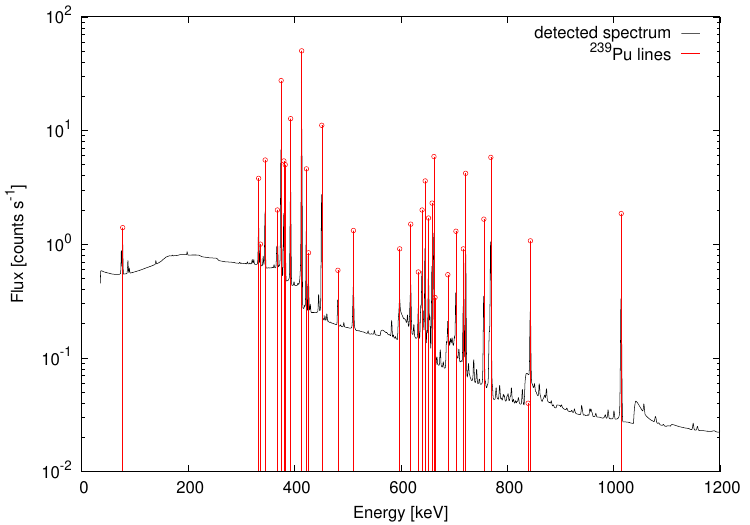}
    \end{tabular}
    \caption{General problem formulation. Left: an illustration of an X-ray radiograph of a scene with image segmentation.
    Middle: spherical object of interest with multiple shielding.
    Right: high-resolution spectrum and lines}
    \label{fig:problem-formulation}
\end{figure*}

The general problem that we seek to examine is one in which a collection of objects are hidden within a scene. 
Their positions, orientations, and most importantly, materials of potential concern, are interrogated with penetrating radiation (see Fig.~\ref{fig:problem-formulation}).
As an initial step, objects are scanned with X-rays, revealing their positions and geometries, possibly with contour finding algorithms obtained from radiographs, from various angles if needed.
We assume here that the geometry or the scene can be recovered with sufficient accuracy using tomographic and segmentation algorithms.
In the problem we are considering, one of the objects contains a radioactive gamma-ray emitting material, such as $^{235}$U or $^{239}$Pu, possibly surrounded by various shielding shells.
We seek to identify the shielding materials by acquiring highly resolved energy spectrum of the intrinsic radiation from a gamma emitting source after it traversed the shielding materials. 
A spectrum with energy resolution necessary to resolve individual lines ($\approx1\%$) can be taken using e.g. high purity germanium (HPGe) detectors~\cite{knoll2010}.

The workflow is illustrated in Fig.~\ref{fig:problem-formulation}, where a spherical object of unknown nature has been revealed as having an onion-like structure of several nested shells.
Figure~\ref{fig:problem-formulation}b shows an example composition of this spherical object, with the innermost shell containing $^{239}$Pu isotope.
Gamma radiation emitted by this radioactive source is attenuated by shielding materials, and the simulated spectrum is shown with a black solid line in Fig.~\ref{fig:problem-formulation}c.
Red lines correspond to the net detector events under each photopeak of $^{239}$Pu to be used for identification of the attenuating materials.

\section{Statistical model}
\label{sec:statmodel}

We perform material identification by assuming a discrete finite set of candidate materials and accompanying geometric measurements for material interfaces.
The procedure consists of computing the trial spectra and comparing them to the test spectrum in an exhaustive manner.
Our prior knowledge consists of the possible materials as well as the assumption that all material combinations are equally likely.
Here, we describe a statistical model for the comparison function to allow for an efficient material identification based on this prior.

The prior of known possible materials and interfaces considerably simplifies the problem.
Firstly, material cross-sections, incorporating all material-specific peculiarities such as recombination edges, are provided and can be taken into account when computing the spectra.
Secondly, there is no need to construct and solve a system of equations for an effective $Z$, as is usually done in dual-energy tomographic material reconstruction.
Finally, instead of working with a full detector spectrum that might feature multiple contaminants irrelevant to the problem, including background continuum, scattering, finite line width and so on, we can extract net events detected under the known line photopeaks and compare only a set of discrete line intensities.
Consequently, a synthetic or experimental spectrum is first processed to extract a set of well-defined peaks.  
This processing includes background subtraction, removal of the continuum, line profile fitting, and accounting for the energy-dependent detector response.
Since the intrinsic line spectra of radioactive isotopes, such as uranium or plutonium, are known and well-studied~\cite{sampson1982,apostol2016,brown2018}, we can use it to calculate the uncollided line flux with added attenuation for each specific trial shielding configuration (\emph{uncollided} as in the total flux emerging from the object before it enters the detector, integrated over the solid angle, in counts per second).
We compute the attenuated line spectrum for all trial material combinations and evaluate the match with the test spectrum using a loss function as defined below.
To enable computational efficiency, an efficient algorithm for calculating the trial spectra is required, as this will allow for a sufficiently fast brute-force combinatorial search.
Accordingly, we have developed a semi-analytic expression for spherically-symmetric and axisymmetric geometries, that only involves a single 1D integration and, as such, is computationally efficient (see Section~\ref{sec:analytic}).

\begin{figure}[!htbp]
    \includegraphics[width=0.95\columnwidth]{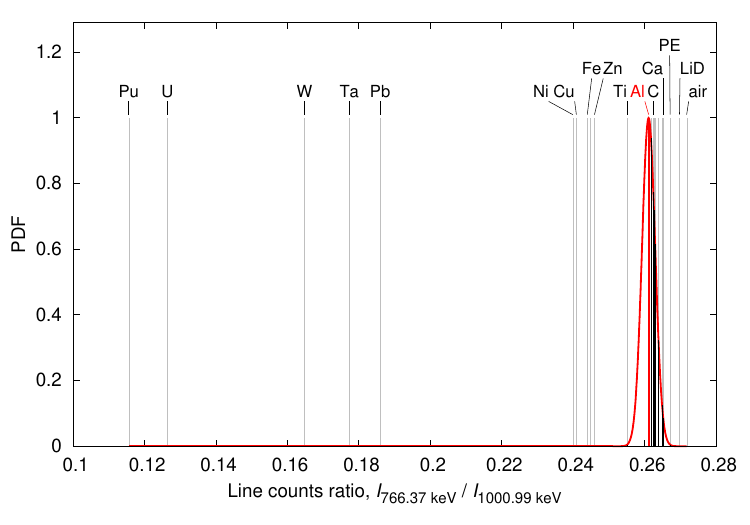}\\
    \includegraphics[width=0.95\columnwidth]{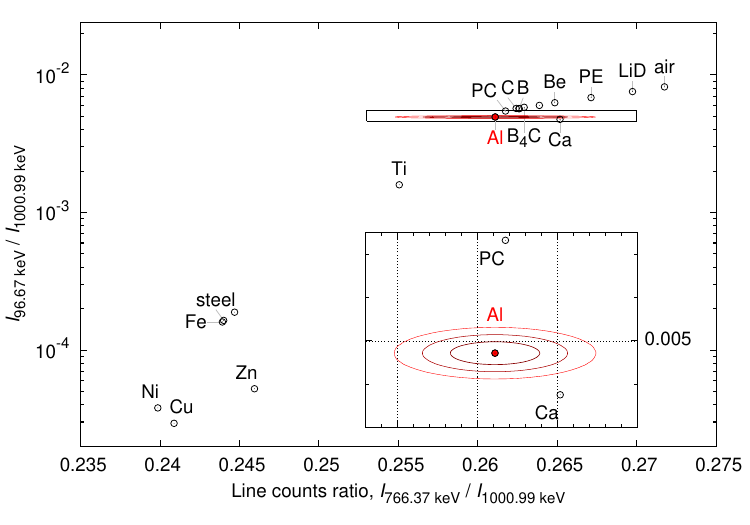}
    \caption{Top: Distribution of the ratio of counts in two strongest 
    lines of $^{238}$U, at the energies $E_1 = 1000.99$~keV and 
    $E_2=766.37$~keV, for a range of various shielding materials. 
    The red curve indicates normal PDF of the true
    value of the line counts ratio for Al, with standard deviation
    $\sigma=0.01$, or 1\% (an error corresponding to about 10000
    counts in both lines).
    Bottom: 2D distribution of relative line counts, for 
    the $E_2/E_1$ ratio (horizontal axis) vs the $E_3/E_1$ ratio
    (vertical axis), where $E_3=94.67$~keV.
    The latter ratio is plotted on a log scale.
    A sequence of highly eccentric nested ovals centered around the
    ratio for Al represents PDF in 2D space of two line ratios, 
    with the same relative standard deviation $\sigma=1\%$, at the
    levels of 0.003, 0.05, and 0.32.
    The levels correspond to the standard 1$\sigma$, 2$\sigma$, and 
    3$\sigma$, confidence intervals.
    Materials include polyethylene (PE), polycarbonate (PC), air, 
    boron carbonate (B$_4$C), steel, lithium-deuterium (LiD) compound,
    and elemental materials labelled by their usual chemical notation.
    } 
    \label{fig:twoline-ratio}
\end{figure}

As an illustration of the principle of our method,
we now consider material identification using two and three lines.
For two lines, the ratio of their counts depends on the material and could be sufficient~\footnote{Material ID could also be performed even with a single line, but in this case the detector needs to be calibrated with a known material, which is in some sense equivalent to having at least two lines.},
as long as different possibilities produce well-separated ratios (for example, in the case of a small number of distinct materials).  
Indeed, consider two lines that are measured to have the ratio of line counts 
$q_{\rm m}:=\mathcal{I}_1/\mathcal{I}_2$, with a measurement uncertainty $\sigma$ (here index $m$ stands for ``measured'').
In the case of high line counts ($>10$), we can safely assume that the true value of $q$ is normally distributed around the measured one: $q\sim\mathcal{N}[q_{\rm m}, \sigma^2]$ with the standard deviation $\sigma$.
Let $k$ be the index for one of the $K$ possible material combinations, 
$k \in 1, 2, \dots, K$,
and $q_k$ the line ratio computed for this combination. 
If we ignore the uncertainty in estimating $q_k$, 
we can claim that the probability of the combination $k$ to match the line ratio $q_m$, measured with variance $\sigma^2$, is estimated as $\mathcal{N}[q_m,\sigma^2](q_k)$.

This approach is illustrated in Figure~\ref{fig:twoline-ratio}, top panel, which shows a PDF centered around a measurement for Al shielding with a relative uncertainty $\sigma=1\%$, on a backdrop of the ratios of counts for various materials.
In this example, the source is a spherical shell of depleted uranium (DU) with inner radius of 5$\sfrac{3}{4}$'' (14.609~cm) and a thickness of $\sfrac14$'' (0.635~cm), and the shielding is a $\sfrac12$''-thick sphere of Al.
We use the two strongest $^{238}$U lines, $E_1 = 1000.99$~keV and $E_2=766.37$~keV.
We can make statements about how well various materials fit the experimental measurement based on the values of this PDF.
As can be seen from the figure, while majority of materials can be easily excluded with high confidence (such as heavy metals, as well as Fe, Ni, Zn, Cu, Ni), several  materials have line ratio well within the margin of error from Al.
For instance, C has practically identical line ratio: ${q_{2/1,{\rm C}} = 0.2624}$ vs ${q_{2/1,{\rm Al}}} =0.2611$ for Al.

Including more than two lines can help increase the discriminating power of the method.
The bottom panel of the Figure~\ref{fig:twoline-ratio} shows the same normal PDF but in 2D space of two line ratios, $q_{2/1}$ and $q_{3/1}$, with the same relative standard deviation of $\sigma=1$\% in both dimensions.
The third line is taken at low energy, $E_3=94.67$~keV.
Here, only a single candidate (Al) falls inside the 3$\sigma$ confidence region (delineated by a very elongated red oval).
The rest of the materials can be excluded at the 3$\sigma$ confidence level.
At the same time, using only the other dimension, i.e. ratio of counts $q_{3/1}$, would not have excluded Ca: this material has a very similar ratio ($q_{3/1, {\rm Ca}} = 0.00475$ against $q_{3/1, {\rm Al}} = 0.00494$ for Al).
Heavy elements, specifically U, Pu, Ta, Pb, and W, have much higher scattering and absorption at $E_3$, such that the lines at this energy are completely absorbed. 
For this reason, these five materials are not shown at the right panel.
This illustrates that having a loss function with more than two lines brings a significant advantage to discriminating power of the method.

Proper choice of the loss function is crucial for reliable discrimination of materials, especially in the presence of noise.
The loss function should be based on the underlying statistical model that properly reflects both the range of possibilities, and our prior knowledge of the measured object.
Specifically for the problem at hand, the prior knowledge consists of us having used a standard procedure for radiographic feature extraction to measure the exact dimensions of the object, its distance to the detector, and the radii of all the shells.
If we assume that in the case of multiple ($L$) measured lines their real count rates are distributed independently around their respective measured rates, and each distribution can be represented by a Gaussian, we can use the following bilinear form for the loss function:
\begin{align}
    \mathcal{L}[\{\mathcal{I}'_\ell, \mathcal{I}_\ell\}]
    :=
    \frac{1}{L}\sum_{\ell=1}^{L} 
                    \frac{(\mathcal{I}'_\ell - \mathcal{I}_\ell)^2}
                         {\sigma_\ell^2},
    \label{eq:lossfunction}
\end{align}
where $\{\mathcal{I}'_\ell\}_{\ell=1}^{L} (\{\mathcal{I}_\ell\}_{\ell=1}^{L})$ is the vector of trial (test) count rates,
and $\{\mathcal{\sigma}_\ell\}_{\ell=1}^{L}$ is the vector of standard deviations for the distribution in each of the lines.
This loss function is a (negative) logarithm of the assumed $L$-dimensional PDF, which in turn is a product of $L$ independent 1D normal distributions with standard deviations as prescribed above.

A point to emphasize is that if real measurement uncertainties were used for the standard deviations $\sigma_\ell$ in (\ref{eq:lossfunction}), this would be suboptimal for discriminating between different materials. 
This is attributed to the fact that the count rates for different lines range over a few orders of magnitude, such that the strongest lines will have a few orders of magnitude higher line counts, and correspondingly, orders of magnitude smaller uncertainties, which will give them much higher relative weight in the loss function (\ref{eq:lossfunction}) above.
Consequently, it is more appropriate to assign each strong line an equal \emph{relative} weight: $\bar{\sigma}\mathcal{I}_\ell$, with some predefined threshold relative uncertainty $\bar\sigma$.
This results in the following loss function, tailored for line pattern matching: 
\begin{align}
    \mathcal{L}[\{\mathcal{I}'_\ell, \mathcal{I}_\ell\}]
    &:=
    \frac{1}{L_{\bar\sigma}\bar\sigma^2}
    \sum_{\sigma_\ell\le\bar\sigma \mathcal{I}_\ell } 
          \left(\frac{\mathcal{I}'_\ell}{\mathcal{I}_\ell} - 1\right)^2 \nonumber\\
    &+
    \frac{1}{L-L_{\bar\sigma}}
    \sum_{\sigma_\ell>\bar\sigma \mathcal{I}_\ell } 
                    \frac{(\mathcal{I}'_\ell - \mathcal{I}_\ell)^2}
                         {\sigma_\ell^2},
    \label{eq:pmlossfunction}
\end{align}
where $L_{\bar\sigma}$ is the number of ``strong'' lines with a relative measurement uncertainty less than $\bar\sigma$. 
In this loss function, the weight of the strong lines is curtailed, while the weak lines contribute with smaller weights $1/\sigma_\ell^2$, corresponding to their higher measurement uncertainties $\sigma_\ell>\bar\sigma\mathcal{I}_\ell$.

\section{Analytic expressions for uncollided flux}
\label{sec:analytic}

\begin{figure}[!htbp]
    \centering
    \includegraphics[width=0.75\columnwidth]{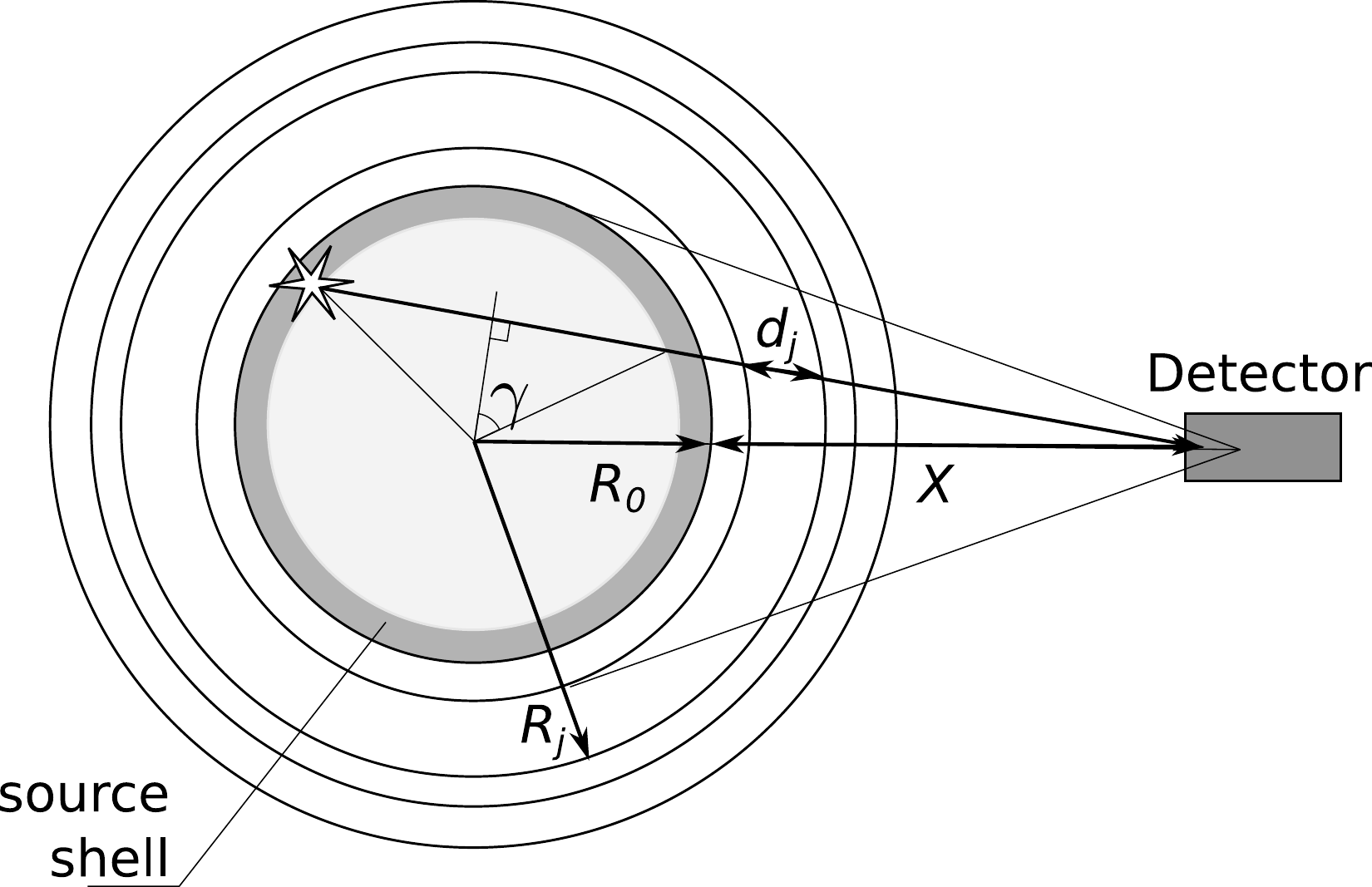}
    \caption{Sketch illustrating integration geometry.}
    \label{fig:integration-sketch-gen}
\end{figure}
For the specific case of the concentric spherical and cylindrical configurations with arbitrary number of concentric shells, general expressions for uncollided flux are derived below.
We assume that each shell is composed of a single uniform material, and that one of the shells is a radioactive source with gamma-emission (see Figure~\ref{fig:integration-sketch-gen}). 
The gamma-emission is then observed from a distance $X$ to the edge of the radiating shell with a detector (see Fig.~\ref{fig:integration-sketch}).
Ignoring in-line scattering and assuming that every shell provides an energy-dependent profile of attenuation $\mu_k(E)$, dictated by the constituting material, we derive the following formula for the uncollided flux, F, from a spherical shielded source (see \ref{app:analytic-deriv} for a detailed derivation):
\begin{align}
    F &= \frac{S_v}{2\mu_0(1+\xi)^2} \nonumber\\
     &\times\int_0^1 x\, dx\left(1 - e^{-2\mu_0 R_0 x}\right)
     \exp{\left\{-\sum_{j=1}^{n}\mu_j d_j(x)\right\}},
\end{align}
where $S_v$ is the gamma-ray source intensity (rate of photons produced per unit volume), $\mu_k$ for $k=0,1,\dots,n$ are the attenuation coefficients for 0$^{\rm th}$ central sphere and 1$^{\rm st}$, ..., n$^{\rm th}$ shells, respectively, $d_j(x)$ are geometric lengths of the rays passing through the j$^{\rm th}$ shell, and $\xi \equiv X_0/R_0$ is the ratio of the distance to the source $X_0$, and the radius of the source sphere $R_0$.
\ref{app:analytic-deriv} gives more details on this notation and the derivations of the uncollided flux for additional configurations of interest, such as when the source is not in the middle, or for cylindrical geomtery.

\section{Experiments and results}
\label{sec:results}
We put the utility of the statistical model and derived analytic approximations to the test by attempting to identify materials using simulated and experimental spectra.
We do this by comparing our synthetic spectra to those generated by a ray-tracing code GADRAS (Gamma Detector Response and Analysis Software \cite{mitchell2009}), as well as to experimental data obtained with a LANL~S germanium HPGe detector.

We examine two types of test objects: 
(a) spherical depleted uranium (DU) shells of varying thickness, covered with Al shells of varying thickness; 
(b) a standard beryllium-reflected plutonium (BeRP) ball~\cite{mattingly2009}. 
DU shells are nested shells each 1/4'' thick, with the outer shell being 12'' in diameter.
The accompanying shielding includes 1/2'' or 1'' thick Al shells.
We refer the reader to a technical report at LANL for detailed description of the experimental setup with DU shells~\cite{stults2020}.
The BeRP ball comes with sets of shielding: 1''-thick steel shell, 1'' thick Ni shell, and 1'' shell made of a mock high explosive (MHE) shell.
The latter is a compound material consisting of 0.0275 H, 0.1681 C, 0.3601 N, 0.4305 O, 0.0031 P, and 0.0107 Cl by mass, with a density of 1.84~g/cm$^{-3}$.
The BeRP ball is made of $\alpha$-Pu (mainly $^{239}$Pu isotope) with a radius of 3.794~cm (about 1.5''), covered with a thin layer of steel alloy (1/3~mm thin), with a 1/3~mm air gap between the surface of Pu and the alloy.
In both setups, the detector is located approximately 1~m from the test sphere, plus elevation, and the collection time in all single-run experiments is 300~s.
Table~\ref{tab:experiments} lists the range of experiments that were conducted.

\begin{table*}
\centering
\caption{Summary of experimental setups. Columns are: the name of the setup, dominant gamma-emitting
isotope, its fraction by mass, radii of the shells, horizontal distance to the detector, 
height above the table, and the collection time. In the last column, ``300x6" means that six measurements were
performed, each taking counts for 300~s, and the results were summed together~\cite{stults2020}.}
\resizebox{\textwidth}{!}{%
\begin{tabular}{l|ccccccc}
\hline\hline
Setup & Dominant & Mass     & Shell radii & Shell areal          & Distance & Height   & Collection  \\
      & isotope  & fraction &     [cm]    & masses [g cm$^{-2}$] & $D$ [cm] & $H$ [cm] & time $\Delta t$ [s]\\
\hline
DU shell:       &           &         &                   &       &     &     &         \\
1/4" + 1/2" Al  & $^{238}$U & 99.7\%  & 14.61,15.24,16.51 & 3.429 & 100 & 12  & 300 x 6 \\
1/4" + 1" Al    & $^{238}$U & 99.7\%  & 14.61,15.24,17.78 & 6.858 & 100 & 12  & 300 x 6 \\
1/2" + 1/2" Al  & $^{238}$U & 99.7\%  & 13.97,15.24,16.51 & 3.429 & 100 & 12  & 300 x 6 \\
1/2" + 1" Al    & $^{238}$U & 99.7\%  & 13.97,15.24,17.78 & 6.858 & 100 & 12  & 300 x 6 \\
3/4" + 1/2" Al  & $^{238}$U & 99.7\%  & 13.34,15.24,16.51 & 3.429 & 100 & 12  & 300 x 6 \\
3/4" + 1" Al    & $^{238}$U & 99.7\%  & 13.34,15.24,17.78 & 6.858 & 100 & 12  & 300 x 6 \\
\hline
BeRP ball +      &           &         &                   &       &     &     &         \\
1" Steel         & $^{239}$Pu & 93.6\% & 3.86,6.40       & 19.926       & 100 & 14   & 300 \\
1" Ni            & $^{239}$Pu & 93.6\% & 3.86,6.40       & 22.611       & 100 & 14   & 300 \\
1" Steel + 2" PE & $^{239}$Pu & 93.6\% & 3.86,6.40,11.48 & 19.926,4.826 & 100 & 13   & 300 \\
1" Ni + 1" MHE   & $^{239}$Pu & 93.6\% & 3.86,6.40,8.94  & 22.611,4.674 & 100 & 17.3 & 300 \\
1" Ni + 2" MHE   & $^{239}$Pu & 93.6\% & 3.86,6.40,11.48 & 22.611,9.347 & 100 & 12.3 & 300 \\
1" Ni + 3" MHE   & $^{239}$Pu & 93.6\% & 3.86,6.40,14.02 & 22.611,14.021& 100 & 12.3 & 300 \\
\hline\hline
\end{tabular}
}
\label{tab:experiments}
\end{table*}

\subsection{Experimental preprocessing}
In an experiment, several sources contribute to confusion: finite line width, neighboring lines, the continuum (including scattered radiation), Poisson noise in the line itself, environmental background, and detector sensitivity.  
Finally, the most important source of confusion arises due to model mismatch between the forward model and the data, and is attributed to imperfect information about the material.
This includes cross sections, composition affecting cross sections, material density, and assumed line intensities.
It should be noted that this source of confusion not only exists when comparing our forward model with experimental data but also when we compare with GADRAS simulations as the ray tracing, cross sections, and interpolation schemes are different.  

Notwithstanding the model mismatch, GADRAS allows for a systematic examination of individual sources of confusion and their influence on the discriminating power of our method.
Previously, GADRAS demonstrated excellent results in preducing realistic estimates of the full gamma spectrum, including full-energy peaks, Compton continua and other features~\cite{rawool-sullivan2012a,rawool-sullivan2012b}.
The measured line flux at the detector for a line $\ell$, in counts per unit time, can be written as:
\begin{align}
    \mathcal{F}_\ell &= \frac{\mathcal{A}\ g(E_\ell)}{4\pi D^2\Delta t} S_v(E_\ell)
    \int_V dV \exp{\left(-\sum_{n\in\mathbf{x}} \mu_n[E_\ell] d_n\right)}\nonumber\\
    &+ \mathcal{B}(E_\ell),
    \label{eq:detected-flux}
\end{align}
where 
 $\mathcal{A}$ is the collection area of the detector,
 $g(E_\ell)$ is the energy-dependent detector sensitivity, 
 $D$ is the distance to the source (see Table~\ref{tab:experiments}),
 $\Delta t$ is the collection time,
 $S_v(E_\ell)$ is the volume emissivity of the radiating source at energy $E_\ell$, 
 and $\mu_n[E_\ell]$ is the attenuation coefficient of the $n$-th material.
Here, the integral is taken over the volume of the source shell, and summation is over all the materials ($n\in\mathbf{x}$) along the path from the integration volume to the detector.
Quantities $d_n$ represent lengths of the segments traversed within each shell (see e.g. Figure~\ref{fig:integration-sketch-gen} and Equation~\ref{eq:dees} in the Appendix for the case of spherical geometry).
The last term on the right, $\mathcal{B}(E_\ell)$ includes environmental background, continuum, scattering from the object, and the Poisson noise, with the latter reflecting the fact that photons are emitted at random times and therefore the detected counts obey Poisson statistic.

In our approach, we identify peaks in either experimental or synthetic spectra, and part of this procedure includes assumptions for the fitted continuum background term $\mathcal{B}(E_\ell)$.
The peak finding algorithm therefore efficiently removes the continuum, only leaving the first term on the right-hand side of Eq.~(\ref{eq:detected-flux}).
The detector efficiency, $g(E_\ell)$, is a known function of energy for the given detector (LANL S) is computed theoretically and verified experimentally~\cite{homan2010}.
After dividing over by the detector efficiency, we are left with the experimental estimate of the peak area, or the uncollided flux in the corresponding line.
This can then be compared to the analytic estimates for matching the unknown materials.

\begin{figure}[!htbp]
    \centering
    \begin{tabular}{c}
    \includegraphics[width=\columnwidth]{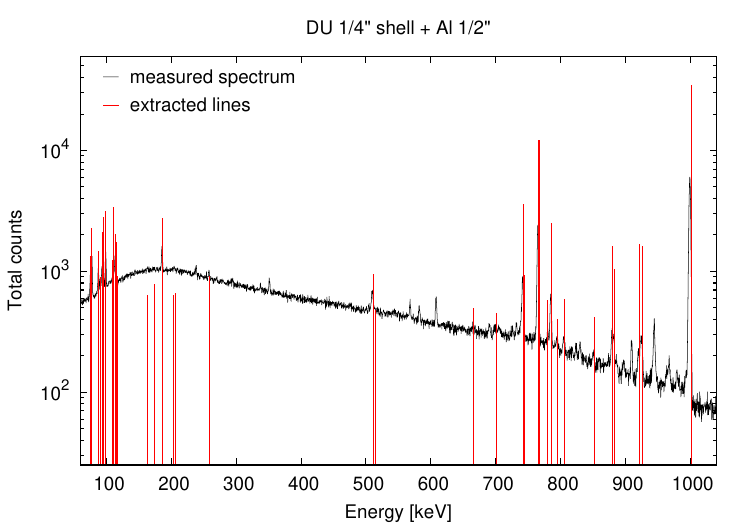} \\
    \includegraphics[width=\columnwidth]{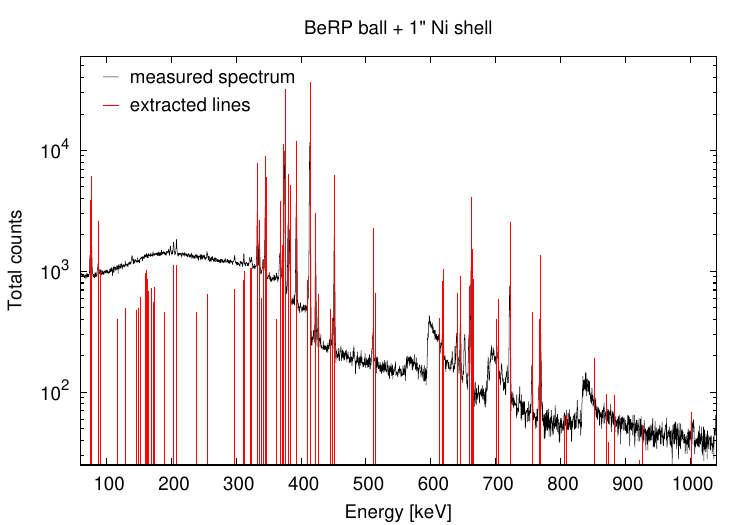}
    \end{tabular}
    \caption{Experimental spectra for 1/4'' DU shell with 1/2'' Al shielding (top), and BeRP ball with 1'' Ni shielding (bottom).
    The red lines indicate integrated peak area counts extracted with PeakEasy (see main text for details).
    } 
    \label{fig:spectrum-exp}
\end{figure}

Figure~\ref{fig:spectrum-exp} shows an example measured spectrum of DU 1/4" shell with 1/2" Al shielding (top), and BeRP ball with 1'' Ni shielding (bottom).
As expected, the spectrum exhibits Poisson noise that appears to be stronger in the lower intensity regions.
We identify several lines with 2$\sigma$-significance, and record their peak energies and intensities, by fitting line profiles over the continuum background.
This is performed using the standard line-fitting algorithms, as provided by PeakEasy software~\cite{rooney2015}.
From the energies of the strongest lines, it is trivial to infer the dominant isotope: $^{238}$U for the DU shells, and $^{239}$Pu for the BeRP ball cases.
The extracted lines are shown in red.

For the dominant isotope, the list of generated gamma-lines with their energies and intensities is well known~\cite{brown2018}.
With this list and a trial material configuration, the ray-tracing algorithms in GADRAS can be used to calculate the corresponding uncollided line flux.
GADRAS also computes the lines due to other isotopes, as well as neutron capture and inelastic scattering lines from the shielding material.
Although these lines can also be potentially utilized for material identification, this is much harder, as they are dependent on subdominant isotopes, complex nonlinear neutron transport, or photon scattering interactions, which we do not intend to concern ourselves with in this work.
In summary, we will only use the lines listed in standard nuclear databases for a single dominant isotope~\cite{brown2018}, and calculate attenuation due to a specific trial material combination at the corresponding energy.

\begin{figure}[!htbp]
    \centering
    \begin{tabular}{c}
    \includegraphics[width=\columnwidth]{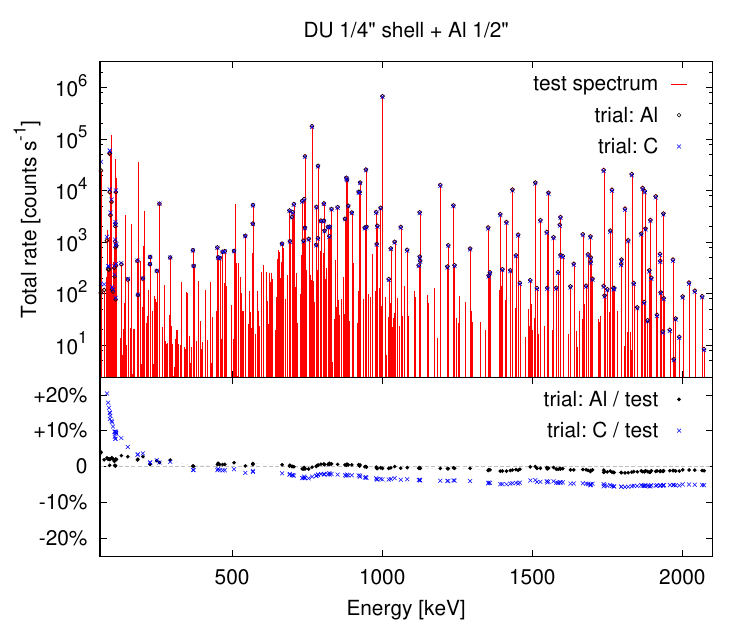} \\
    \includegraphics[width=\columnwidth]{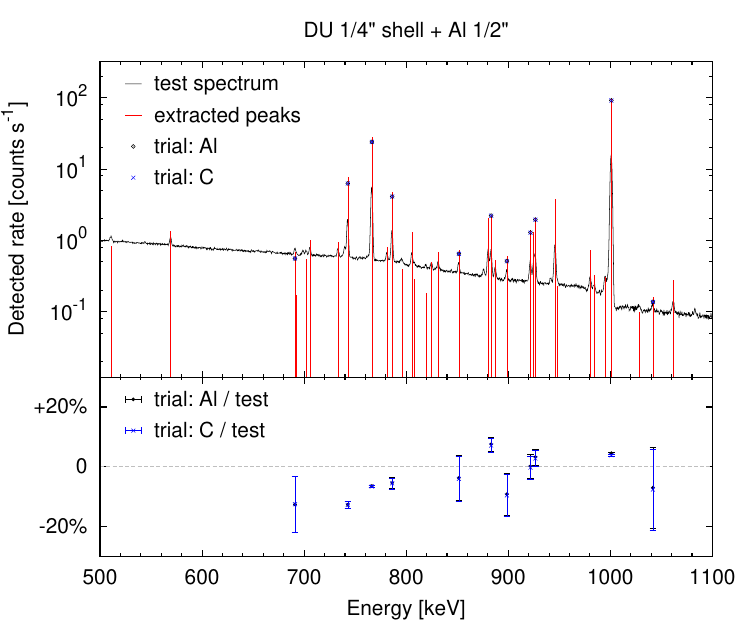}
    \end{tabular}
    \caption{Uncollided total (top) and simulated detected (bottom) count rate from a spherical 
    1/4'' DU shell, shielded by 1/2'' of Al. 
    The red lines represent line fluxes as reported by GADRAS for this object, either 
    uncollided (top), or from integrating the peak areas (bottom).
    Black circles and blue crosses represent our trial analytic estimates for Al and C, respectively.
    The black solid curve on the right shows the continuous spectrum, simulated for a HPGe detector using GADRAS.
    The lower part of each plot zooms on the ratio of trial to test line fluxes (in percentage of difference from 1), with the right panel also showing the uncertainties for each ratio due to the peak fitting.
    It should be noted that some of the lines do not have analytic estimates. 
    This is because they are only computed for the dominant isotope, $^{238}$U, while other isotopes are also present in the DU composition.
    Additional lines can be identified in the spectra, such as 511~keV line visible in both panels, are not directly produced by radioactive decays, but rather other processes such as neutron captures or inelastic scattering on Al (in particular case of 511~keV line, by $e^+ e^-$-annihilation).
    } 
    \label{fig:spectrum-DU}
\end{figure}

Figure~\ref{fig:spectrum-DU}, top panel, shows the uncollided line flux from a spherical 1/4''-thick DU  shell, covered by 1/2''-thick Al shielding. 
The blue crosses and black circles indicate corresponding line fluxes as calculated by our analytic formula, but only for the lines identified as coming from the dominant isotope, $^{238}$U in this case (see 2$^{\rm {nd}}$ column in Table~\ref{tab:experiments} for the dominant isotopes in our setups).
The rest of the lines come either from subdominant isotopes such as $^{235}$U, inelastic scattering, or from neutron captures, in this case on Al.

\begin{figure*}[!htbp]
    \centering
    \begin{tabular}{cc}
      \includegraphics[width=0.45\textwidth]{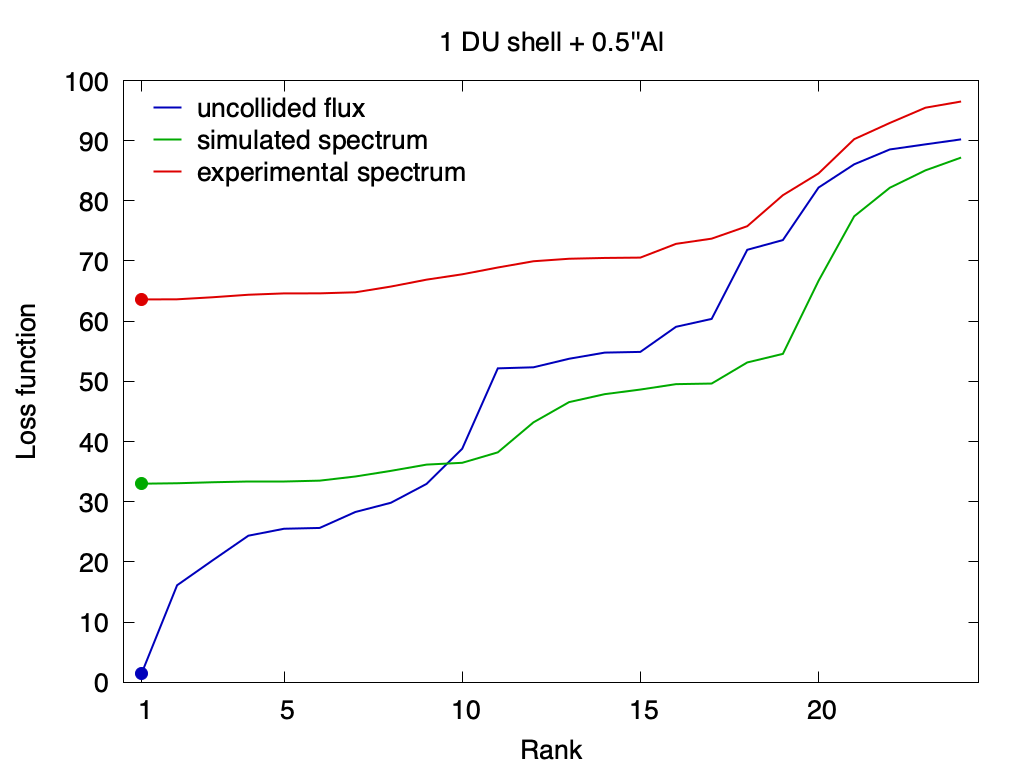} &
      \includegraphics[width=0.45\textwidth]{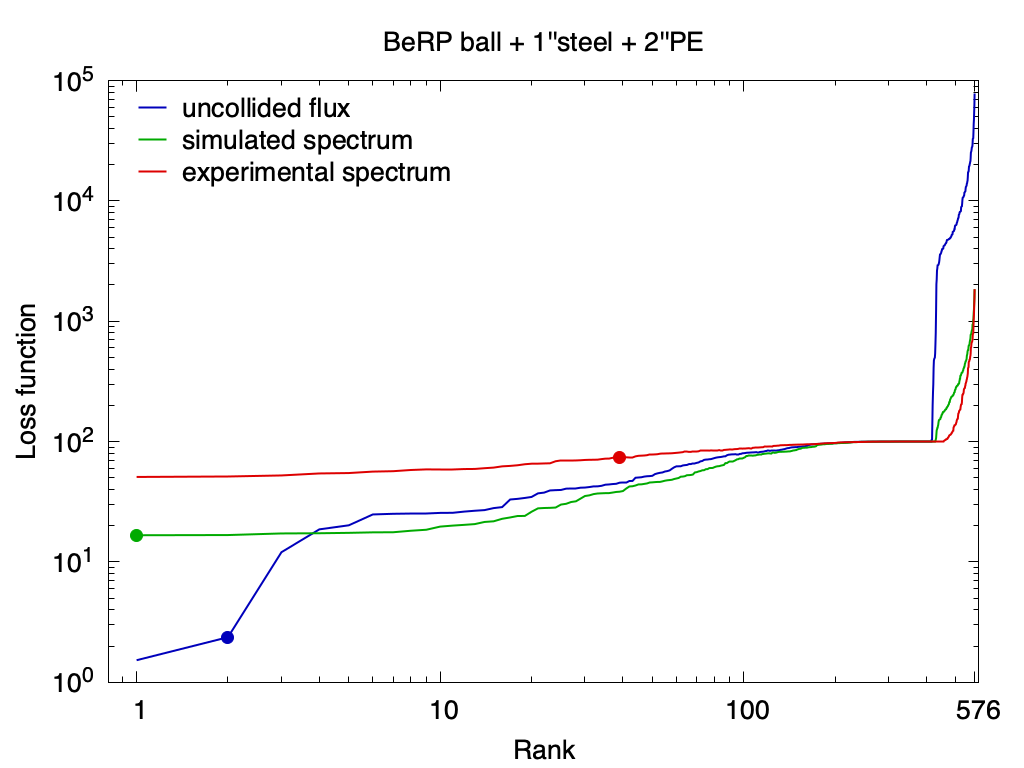}
      \\
      \includegraphics[width=0.45\textwidth]{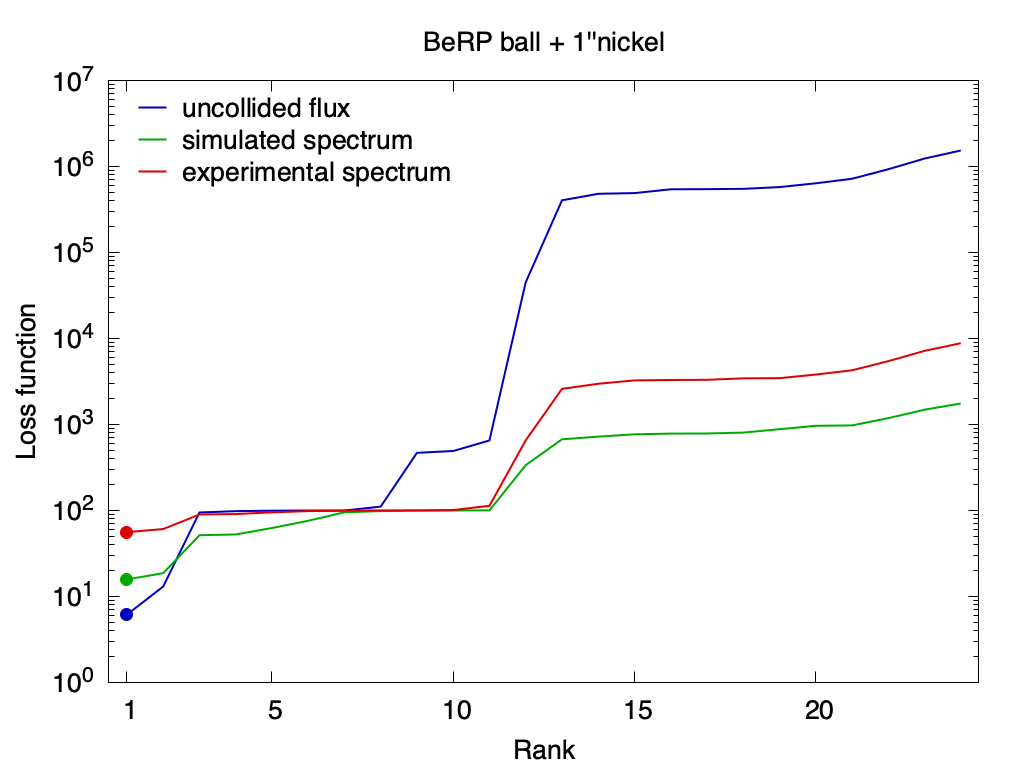} &
      \includegraphics[width=0.45\textwidth]{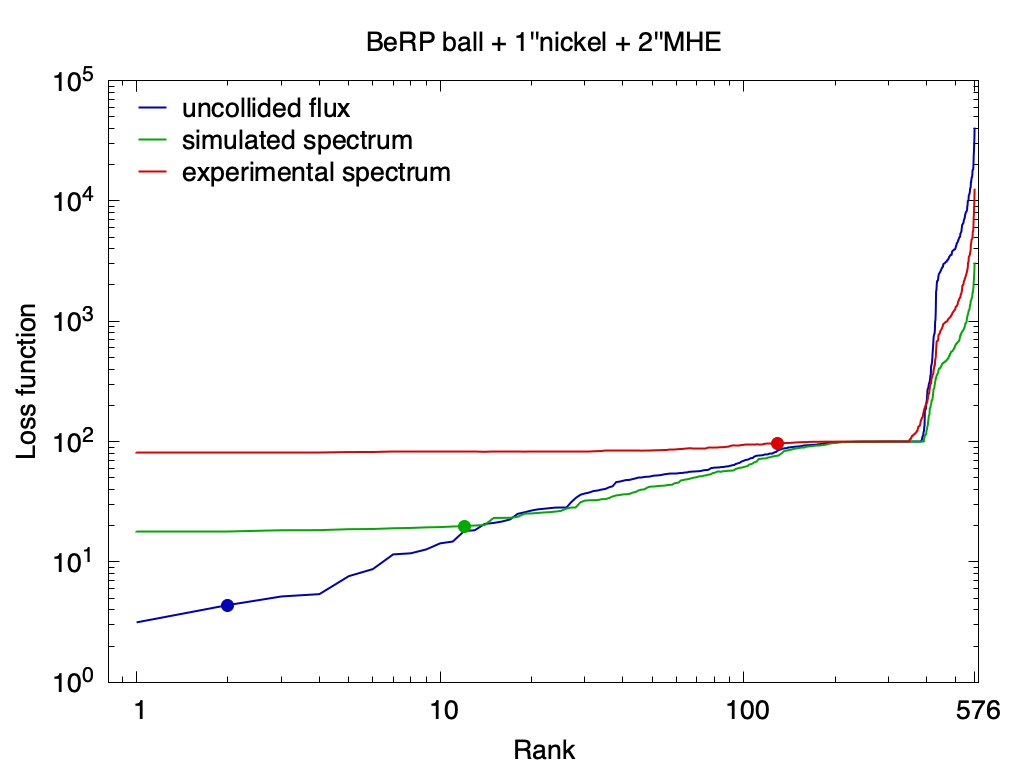}
    \end{tabular}
    \caption{The loss function for a range of single-material (left column) and
    two-material (right column) combinations, ordered by rank, for the uncollided flux 
    (blue), from peaks in simulated LANL~S spectra (green), and from the experiment.
    The dots mark the ground truth trials.
    Notice that the scale is logarithmic in the x-axis for two-material combinations.
    } 
    \label{fig:id_single_double_mat}
\end{figure*}

The bottom panel of the same figure shows a GADRAS-simulated spectrum for this object as it would have been detected by the LANL S detector.
This simulation, however, does not include Poisson noise for individual energy bin counts, since we want to first investigate the influence of continuum background in a deterministic manner.
To extract the lines, we use a standard peak fitting procedure (implemented in GADRAS), where line profiles are fit on top of a local linear background.
Extracted line fluxes are shown as red vertical lines.
As can be seen, the lines are now a subset of lines from the uncollided flux; this is not only because many lines are submerged in the continuum, but also due to the fact that they are not \emph{well-separated} from other lines.
In our analysis, we do not consider lines that are too close to other lines. 
A tolerance of $\delta E = 0.04$~keV is applied when matching known lines and the energies of the peaks found in the line profile fitting algorithm.
Namely, if the line fitting algorithm reports peak energy at $E_{\ell,{\rm fit}}$, it is considered a match with a known line $E_\ell$, if: (a) $|E_{\ell,{\rm fit}}-E_\ell| < \delta E$, and (b) there is only one line satisfying this condition.
Applying these requirements reduces the total number of fitted lines to a much smaller number (on the order of ten) compared to uncollided flux (on the order of a hundred---cf. panels in Fig.~\ref{fig:spectrum-DU}).

\subsection{Results}

We apply a brute-force approach to examine all possible material combinations and compute the corresponding loss function for each of them.
To find the best fit, we sort the loss functions in increasing order and identify the top candidates.
Figure~\ref{fig:id_single_double_mat} shows the loss function for different materials as a function of rank after sorting. 
Every figure shows three cases: (a) uncollided flux (blue lines), (b) simulated spectrum detection with GADRAS, assuming absence of Poisson noise (green lines), and (c) experimental spectral measurements (red lines).
Points on each curve indicate ground truth, i.e. correct material.
In the ideal case, ground truth is identified as having rank 1, with a minimal loss function across all possible material combinations.
We consider 24 different materials, with $24^2=576$ two-material combinations.

As may be seen from the Figure~\ref{fig:id_single_double_mat}, the method applied to lines from the uncollided flux always yields ground truth i.e. material combination rank \#1. 
Moreover, the value of the loss function for rank \#1 combination is noticeably lower than the subsequent ranks.
But for simulated spectra, the loss function curve shows a broad ``plateau'' near minimum, such that the minimum value does not differ significantly from the surrounding following ranks.
In this situation, discrepancies in cross-sections, interpolation, and ray-tracing between GADRAS and our code (model mismatch) cause small differences in the loss function, such that the ground truth value may or may not have the rank \# 1. Furthermore, additional differences may be induced via the continuum subtraction and peak finding algorithm.
This situation becomes even more pronounced for the experimental loss function, where additional sources of uncertainty are contributing to model mismatch and resulting degradation of the ranking.
The loss function exhibits an extended plateau, and the ground truth can be found anywhere on it (see e.g. right panels in Figure~\ref{fig:id_single_double_mat}).

\begin{figure}[!htbp]
    \centering
    \begin{tabular}{c}
      \includegraphics[width=\columnwidth]{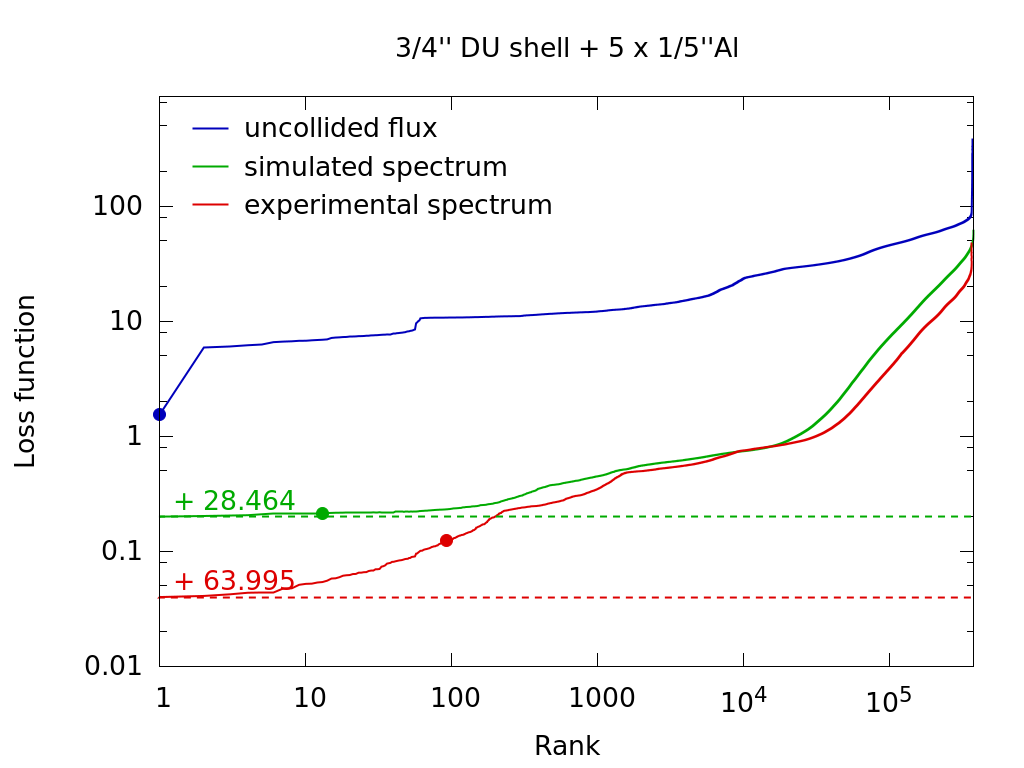} \\
      \includegraphics[width=\columnwidth]{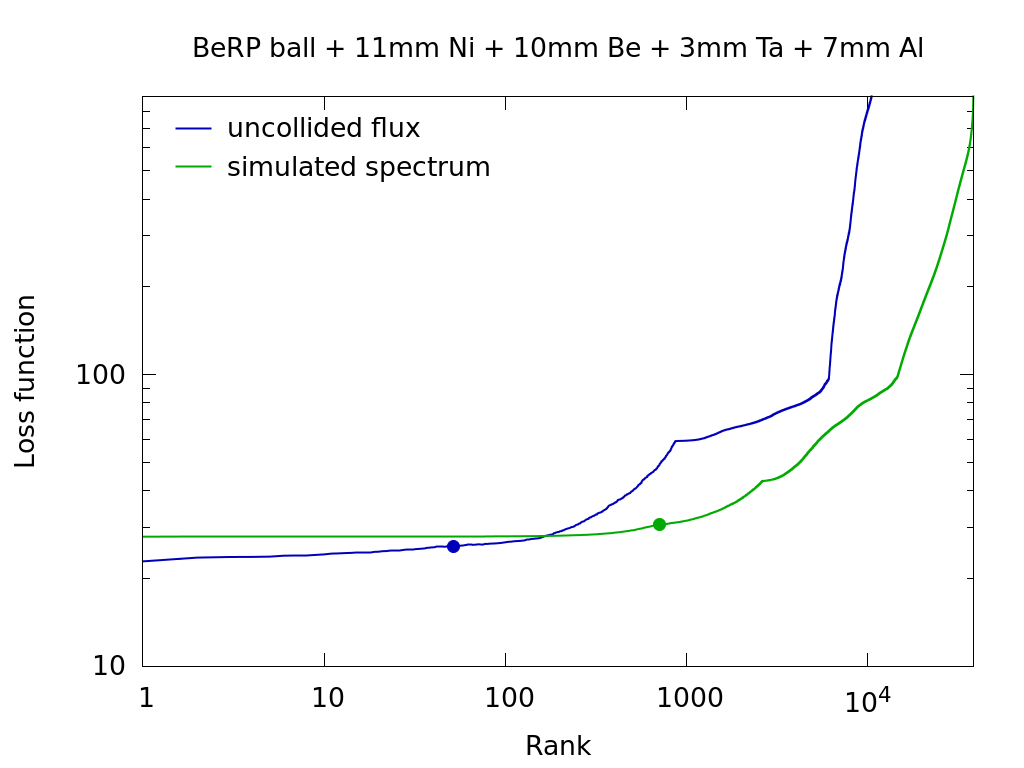}
    \end{tabular}
    \caption{Loss function ordered by rank, used for material identification with 
    5 (top) and 4 (bottom) unknown materials.
    Dots mark location of the ground truth, as in the previous Figure.
    Left: 3/4'' DU shell with 1'' Al shell that is split into five 1/5''-thick subshells, treated as five unknown materials.
    To highlight the behavior of the loss functions for simulated and experimental spectra, we subtracted its values at the rank \#1 and added 0.2 and 0.04, respectively.
    Right: simulation of a BeRP ball with Ni, Be, Ta, and Al shielding shells of varying thickness.
    In this case, there is a noticeable decrease in the performance of the method.
    We attribute this to a model mismatch between GADRAS and our code, including discrepancies in the input data, such as material cross-sections used in both codes.
    } 
    \label{fig:id_DU_1inAl_multiple}
\end{figure}

Figure~\ref{fig:id_DU_1inAl_multiple} presents the loss function behavior for multiple materials, for two cases with (five and four) unknown materials.
In the top panel, a 1'' Al shell around \sfrac{3}{4}'' of DU is artificially split into 5 thinner shells, and we attempt to identify each of these by a brute-force search through $14^5=537\,824$ material combinations.
Here, for the green and the red curves that represent GADRAS-simulated and experimentally measured spectra, respectively, the loss function shows an extended plateau.
To examine the loss function dynamics in greater detail, we plot the loss functions with subtracted values at rank \#1 and added 0.2 and 0.04, resp.
The top panel of Figure~\ref{fig:id_DU_1inAl_multiple} demonstrates that the method successfully identifies correct combination for the uncollided flux, and gives very reasonable results not only for synthetic GADRAS spectrum without Poisson noise, but also for actual experimental values, where the ground truth combination is ranked within the first 100 out of half a million possibilities (in the top 0.019 percent).
It also shows that for the approximately first $100$ combinations, the loss function does not deviate from its lowest value by more than 0.1, demonstrating an extended plateau. 
Finally, this example suggests that the determination of the material boundaries via a radiograph may not even be necessary: it is sufficient to subdivide it into thin shells and optimize for the best-fitting material combination.
That is, in the previous example we simply divided the single aluminum shell into five arbitrary shells, and they were correctly inferred to be made of aluminum.
However, this process cannot be applied for arbitrarily thin shell splitting: first, because computational cost for many subshells grows as a factorial of their number, and secondly, because very thin subshells have little areal mass and cannot be reliably identified.

The bottom panel in Figure~\ref{fig:id_DU_1inAl_multiple} presents a simulated scenario with four shielding shells around BeRP ball, where all materials are different.
Here we only demonstrate the results with the uncollided flux and lines peaks extracted from the synthetic spectrum as experimental data is not available for this configuration. 
In this case it is observed that even in the case of uncollided flux the ground truth is ranked at 52.
This must be due to the accumulated model mismatch between the GADRAS model and our forward one.
Likewise, for the simulated spectrum (green line on this panel) the additional confusion added by the imperfect continuum removal causes the ground truth to approach 700 (out of $14^4=38416$, or in the top 2\% of all candidate material combinations).

Table~\ref{tab:results} provides a comprehensive report of the material identification using our statistical method. 
The first two columns contain the setup and the number of assumed unknown shielding materials.
We examined two kinds of experimental setups: one with DU shells, and another with a BeRP ball in addition to various numerical investigations.
In the experiments, only one or two shielding materials were used, but for the  numerical investigations, we artificially subdivided shielding shells into several subshells (e.g. 1'' Al $\to$ 3x1/3'' Al etc.)
This way, we can re-use experimental data in a more comprehensive manner.
Finally, we also examined two additional numerical cases using a BeRP ball and four layers of shielding: 11\,mm Ni, 10\,mm Be, 3\,mm Ta, and 7\,mm Al for the first case, and the same configuration but with polyethylene (PE) instead of Be for the second case.
Although there is no experimental data available for these two cases, but they serve as a test of the limits of applicability of our method.

The third column in Table~\ref{tab:results}, $\delta\mathcal{L}_{1,2}$, represents the distance, in the sense of Equation~\ref{eq:pmlossfunction}, between the ground truth line spectrum and the most similar other spectrum over all examined configurations:
$\delta\mathcal{L}_{1,2} = \min_{j} \mathcal{L}[\{\mathcal{I}_{\rm GT}\},\{\mathcal{I}_j\}]$.
These spectra are computed using our analytic model for all possible material combinations in a given setup, for the purpose to demonstrate distinguishability of these combinations.
If two material combinations were to produce identical attenuated spectra, the distance $\delta\mathcal{L}_{1,2}$ would have been zero, rendering these combinations indistinguishable.
The third column in Table~\ref{tab:results} shows that this never happens for any of our setups.
The value of $\delta\mathcal{L}_{1,2}$ also provides an estimate of the magnitude of the loss function for unambiguous material identification.
This condition is for the loss function to satisfy $\mathcal{L}_1\ll\mathcal{L}_{1,2}$.
Table~\ref{tab:results} shows that in all cases where the loss function of the top candidate $\mathcal{L}_1$ is significantly smaller than $\mathcal{L}_{1,2}$, this candidate is the ground truth (ranked \#1).  
On the other hand, if this condition is not satisfied, there is a difficulty in performing material identification, i.e. the lower $\delta\mathcal{L}_{1,2}$ compared to $\mathcal{L}_1$, the more difficult  material identification will be in general.

\begin{table*}
\centering
\caption{Summary of the material identification results for experimental and numerically simulated setups.
The first and the second columns contain the setup label and the number of unknown materials to be identified. 
The third column $\delta\mathcal{L}_{1,2}$ is the distance (in the sense of Equation~\ref{eq:pmlossfunction}) between the ground truth spectrum and closest spectrum over all examined configurations.
The next three column sections refer to: 
(1) uncollided flux: GADRAS estimates are used, 
(2) GADRAS-simulated spectra with LANL S detector, and
(3) experimentally measured spectra.
In each of these column sections, the first column shows rank of the ground truth (GT) combination, $\mathcal{L}_1$ shows the loss function of the top combination, and the third column lists either the loss function of the closest-ranking second combination, $\mathcal{L}_2$ (if GT scored first), or the loss function of the GT, $\mathcal{L}_{\rm GT}$.
The value of the loss function for GT is highlighted in bold.}
\resizebox{\textwidth}{!}{%
\begin{tabular}{lc|c|ccc|ccc|ccc}
\hline\hline
 & & & \multicolumn{3}{c|}{uncollided}
     & \multicolumn{3}{c|}{simulated}
     & \multicolumn{3}{c}{experimental}\\
 & & & \multicolumn{3}{c|}{flux} 
     & \multicolumn{3}{c|}{spectra} 
     & \multicolumn{3}{c}{spectra}\\
Setup & \#m & $\delta\mathcal{L}_{1,2}$
     & rank & $\mathcal{L}_1$ & $\mathcal{L}_{2:{\bf GT}}$ 
     & rank & $\mathcal{L}_1$ & $\mathcal{L}_{2:{\bf GT}}$
     & rank & $\mathcal{L}_1$ & $\mathcal{L}_{2:{\bf GT}}$ \\
\hline
DU shell:       &   &       &    &            &        &   &           &        &   &           &     \\
1/4" + 1/2" Al  & 1 & 16.2  &  1 & {\bf 1.40} & 16.2   & 1 & {\bf 32.9}& 33.0   & 1 & {\bf 63.6}& 63.6 \\
1/4" + 1" Al    & 1 & 29.6  &  1 & {\bf 1.84} & 29.9   & 1 & {\bf 28.8}& 29.6   & 1 & {\bf 64.4}& 64.7 \\
1/2" + 1/2" Al  & 1 & 16.1  &  1 & {\bf 1.48} & 15.9   & 1 & {\bf 29.7}& 29.9   & 2 & 64.3 & {\bf 64.3}\\
1/2" + 1" Al    & 1 & 29.4  &  1 & {\bf 1.90} & 29.6   & 1 & {\bf 28.6}& 29.0   & 1 & {\bf 66.5}& 67.2 \\
3/4" + 1/2" Al  & 1 & 16.0  &  1 & {\bf 1.23} & 15.6   & 1 & {\bf 33.0}& 33.1   & 2 & 65.5 & {\bf 67.0}\\
3/4" + 1" Al    & 1 & 29.3  &  1 & {\bf 1.56} & 29.2   & 1 & {\bf 28.7}& 29.4   & 1 & {\bf 64.1}& 64.6 \\
3/4" + 3x1/3" Al& 3 & 8.82  &  1 & {\bf 1.56} & 9.44   & 1 & {\bf 28.7}& 28.8   & 1 & {\bf 64.1}& 64.1 \\
3/4" + 4x1/4" Al& 4 & 7.44  &  1 & {\bf 1.56} & 7.19   & 1 & {\bf 28.7}& 28.7   & 43& 64.1 & {\bf 64.1}\\
3/4" + 5x1/5" Al& 5 & 6.13  &  1 & {\bf 1.56} & 5.90   & 13& 28.7 & {\bf 28.7}  & 92& 64.0 & {\bf 64.1}\\
\hline
BeRP ball +                  &   &       &    &     &        &   &        &    &    &           &      \\
1" Steel         & 1 &  31.1  & 1 &{\bf 1.50} & 2.78  & 2 & 22.8 & {\bf 26.9}  &  1 & {\bf 47.5}& 47.5 \\
1" Ni            & 1 &  11.8  & 1 &{\bf 6.13} & 13.0  & 1 & {\bf 15.7} & 18.6  &  1 & {\bf 55.7}& 60.6 \\
1" Steel + 2" PE & 2 &  1.25  & 1 &{\bf 1.53} & 2.37  & 1 & {\bf 16.6} & 16.7  & 39 & 50.8& {\bf 73.8} \\
1" Ni + 1" MHE   & 2 &  1.83  & 4 &3.37& {\bf 4.33}   & 1 & {\bf 17.6} & 17.7  & 123& 54.1& {\bf 65.4} \\
1" Ni + 2" MHE   & 2 &  4.48  & 2 &3.15& {\bf 4.38}   & 6 & 18.0 & {\bf 19.8}  & 129& 81.7& {\bf 96.7} \\
1" Ni + 3" MHE   & 2 &  1.49  & 1 &{\bf 4.10} & 4.75  & 5 & 13.0 & {\bf 14.1}  & 43 & 65.5& {\bf 69.7} \\
3x1/3" Steel     & 3 &  0.64  & 1 &{\bf 1.50} & 1.79  & 39& 16.6 & {\bf 21.0}  &  1 & {\bf 47.5}& 49.0 \\
3x1/3" Ni        & 3 &  4.30  & 1 &{\bf 6.13} & 6.23  & 1 & {\bf 13.3} & 15.4  &  1 & {\bf 55.7}& 57.0 \\
1" Ni + 2x1" MHE & 3 &  2.29  & 31& 2.14 &{\bf 4.38}  & 79& 17.5 & {\bf 19.5}  & 680& 81.7& {\bf 96.7} \\
4x1/4" Steel     & 4 &  0.48  & 1 &{\bf2.78} & 7.77   &442& 15.0 & {\bf 26.9}  &  1 & {\bf 47.5}& 48.6 \\
4x1/4" Ni        & 4 &  3.20  & 5 & 5.88& {\bf 6.13}  & 1 & {\bf 15.8} & 15.9  &  1 & {\bf 55.7}& 56.7 \\
1" Ni + 3x1" MHE & 4 &  0.89  & 446&12.8& {\bf 13.9}  &559& 12.8 & {\bf 14.1}  &4772& 64.6& {\bf 69.7} \\
2x1/2" Steel + 2x1" PE &4&0.54& 18 &1.25& {\bf 2.37}  & 59& 1.25 & {\bf 2.37}  &7702& 48.6& {\bf 74.4} \\
2x1/2" Ni + 3x1" MHE   &5&0.17& 518&0.79& {\bf 4.10} &1342& 12.8 & {\bf 14.1}  &2680& 66.3& {\bf 69.7} \\
Ni + Be + Ta + Al& 4 &  2.52  & 52 & 22.9 &{\bf 25.8}  &712& 27.8 & {\bf 30.6} & &- \\
Ni + PE + Ta + Al& 4 &  2.29  & 155& 16.4 &{\bf 18.6}  &890& 17.5 & {\bf 22.1} & &- \\
\hline\hline
\end{tabular}
}
\label{tab:results}
\end{table*}

The ``uncollided flux'' columns in the Table~\ref{tab:results} shows the results for material ID using the data GADRAS computes for ground truth uncollided flux.
These data represent total individual line count rates computed with GADRAS ray tracing algorithm, before they enter the detector, integrated over the entire solid angle.
The first column is the rank of the ground truth material combination, the second column is the loss function of the top candidate, and the third column is either the loss function of the second-ranked candidate if the ground truth scored first, or the loss function of the ground truth.
Here, non-zero loss function with our models is solely attributed to model mismatch, which includes differences in assumed cross-sections, interpolation errors between the data points, and truncation error in ray tracing.
As can be observed from the Table, for DU shells tests with the uncollided flux, the ground truth combination always ranks first, even for the cases with 5 materials.
This is perhaps not surprising, because $\mathcal{L}_{\rm GT}\ll\mathcal{L}_{1,2}$ for all these cases.

For the BeRP ball setups with two or more unknown materials, $\mathcal{L}_{\rm GT}\approx\mathcal{L}_{1,2}$ and material ID becomes ambiguous.
In the worst case, for the setup 2x1/2" + 3x1" MHE with 5 materials, $\mathcal{L}_{\rm GT}>\mathcal{L}_{1,2}$, and the GT candidate scores at rank 518 (in the top $600/14^5 \approx 0.1$\%).
Even though this is out of more than half a million candidates, it is not ideal, and demonstrates the limits of our approach in terms of its sensitivity to the model mismatch.
Additionally, for these setups, MHE shells have rather small areal mass (see Table~\ref{tab:experiments}), and produce very weak attenuation, which hinders material identification.

The ``simulated spectrum'' section in the Table~\ref{tab:results} shows the results for material ID using the GADRAS-simulated spectrum with peak area identification.
Poisson noise is not added in these spectra, and the detector response is efficiently factored out for this case; compared to the uncollided flux, finite line width and continuum due to scattering are added.
Addition of the continuum eliminates very weak lines, leaving only on the order of 10 strongest lines for comparison with the model.
This effectively increases the loss function such that $\mathcal{L}_{\rm GT}\approx\mathcal{L}_{1,2}$, but the results for the DU shells are still rather good.
Conversely, the situation is suboptimal for multi-material ID with BeRP ball setups, where GT scores in the first thousand for 4 and 5 materials (still within the top 0.1\%).
We attribute this to the fact that $\mathcal{L}_{\rm GT}\gg\mathcal{L}_{1,2}$ in all of these cases.
More advanced methods of continuum subtraction that take into account peculiarities in the measured spectrum can help improve these results.

Finally, the ``experimental spectra'' section in the Table~\ref{tab:results} lists material ID using the experimental spectra and line peak areas computed from them.
Here, the loss function further increases by a factor of $\approx2$, due to now present Poisson noise but more importantly, due to the noise imparted by imperfect environment background subtraction.
This slightly degrades the result for DU shell material ID, with 4 and 5 materials GT now scoring only in the first hundred (top $0.018\%$).
The single-material ID for BeRP ball setups still show excellent scores, but suboptiomal scores for two and more materials.

The numerical experiments above suggest that, for successful material identification with multiple materials, both the line fluxes as well as the material properties that factor into the analytic calculations need to be measured more accurately to obtain the top ranking.
The bilinear form~(\ref{eq:pmlossfunction}) that we use for our loss function, can also serve as a measure of distance in the flux space.
With the parameter $\bar{\sigma}=10^{-2}$, this distance can be interpreted as an average difference per line flux, in percent.
In particular, for uncollided flux, the loss function $\approx 1-10\%$ in Figure~\ref{fig:id_single_double_mat} is due to a model mismatch between our analytic estimates and GADRAS. 
The resulting difference in line fluxes can be seen on the bottom left panel in Figure~\ref{fig:spectrum-DU} for trial Al case, where the black points deviate from a perfect horizontal line, as would have been the case for matching models.
These model mismatches seem to be mainly coming from the differences in material attenuation coefficients, and in the line branching ratios that factor into our analytic calculation and are usually given with an accuracy of about 1\%.

\begin{figure}[!htbp]
    \centering
      \includegraphics[width=\columnwidth]{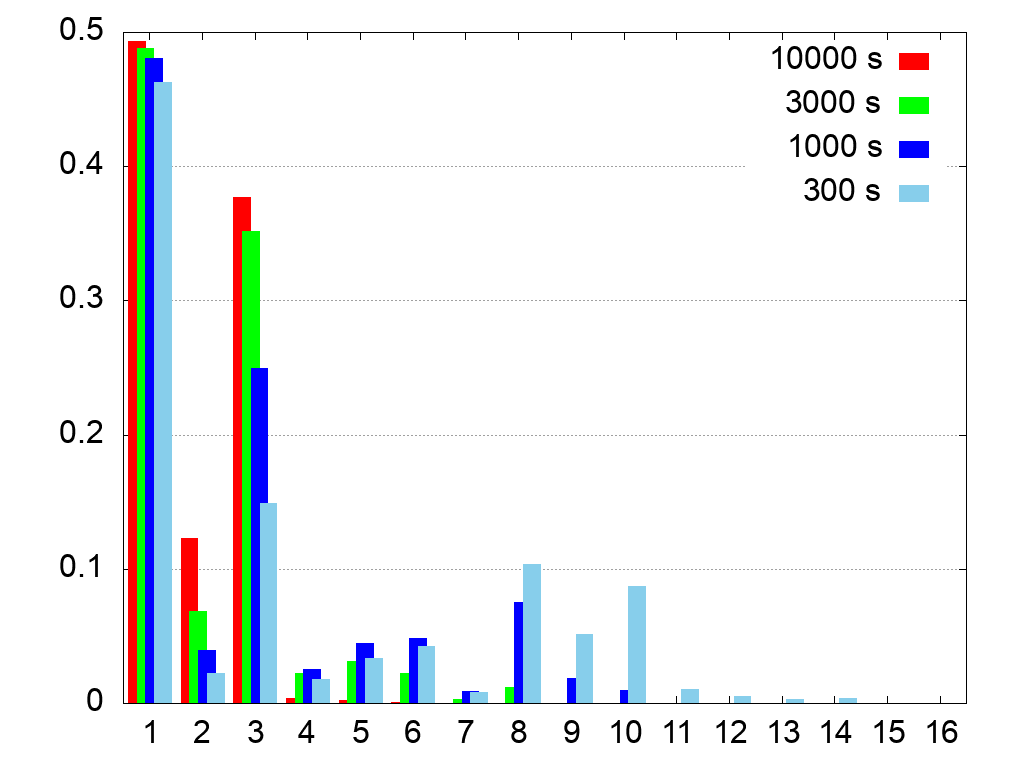}
    \caption{Histogram of the estimated rank probabilities of the ground truth with added Poisson noise, for the BeRP ball + 1'' Steel + 2'' PE configuration, using different collection times, from 300~s to 10000~s.
    Probabilities to get rank higher than 3 decrease with increased collection time.
    } 
    \label{fig:fighting-poisson}
\end{figure}

The quantum nature of light, statistical nature of radioactive decay, and finite collection time lead to inevitability of Poisson noise in the experimental measurements.
Figure~\ref{fig:spectrum-exp} shows that, for collection times that are below one hour, typical photon counts for strongest lines in BeRP ball experiments reach $10^3-10^4$, giving a line count error of $1-3\%$ due to the Poisson noise.
It is therefore impossible to discriminate between material combinations that require sub-percentile average relative line flux sensitivity for unambiguous identification.
However, with increased photon collection time, the results can be improved,
as shown in Figure~\ref{fig:fighting-poisson}, where we demonstrate estimates of the probability for the ground truth to have rank lower (worse) than \#1 due to Poisson noise.
For the lowest collection time, 300~s, there is a non-negligible probability for get ground truth scoring as low as 14. 
With increased collection time, distribution leaves progressively less chance to have ground truth scoring worse than 1.
This demonstrates the range of ranking error that is to be expected due to Poisson statistics.

With Poisson noise or model mismatch present, the loss function for the top-ranking material combinations forms a slowly rising plateau, with no strong preference for any of the candidates.
In potential practical applications of the suggested technique, the extent of the plateau needs to be determined, giving a list of candidate combinations.
All of the combinations in the list are roughly equally likely.
With increased collection time, the extent of the plateau in the loss function shrinks, giving fewer candidates.

\begin{figure*}[!htbp]
    \centering
    \begin{tabular}{ccc}
      \includegraphics[width=0.33\textwidth]{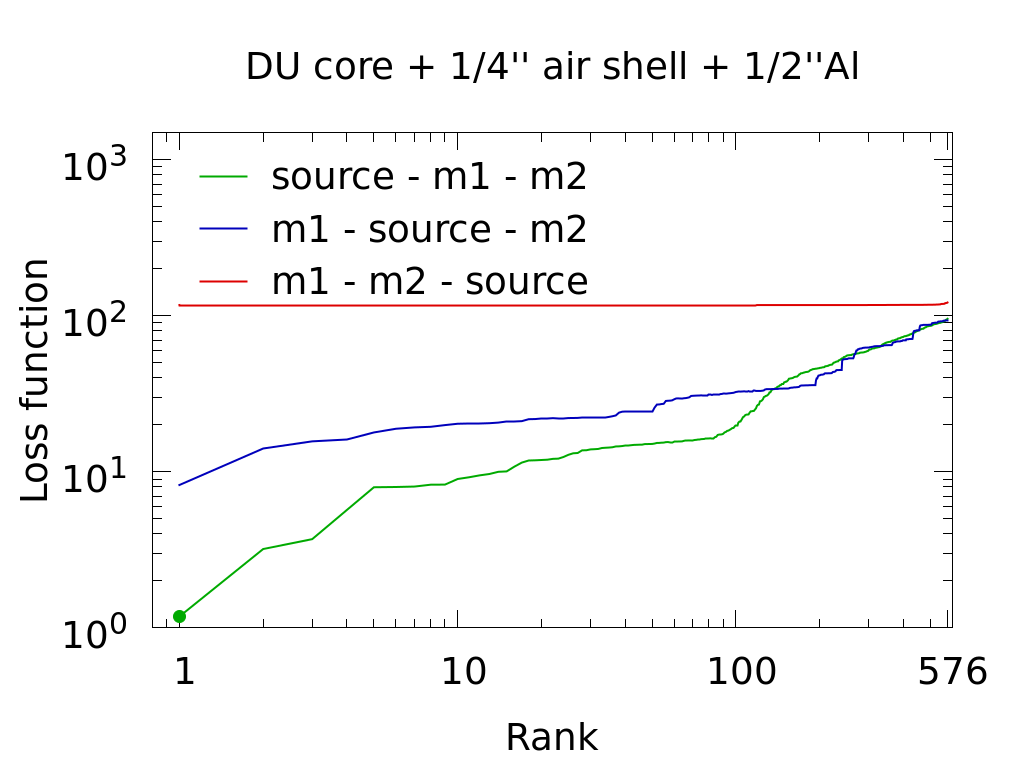} &
      \includegraphics[width=0.33\textwidth]{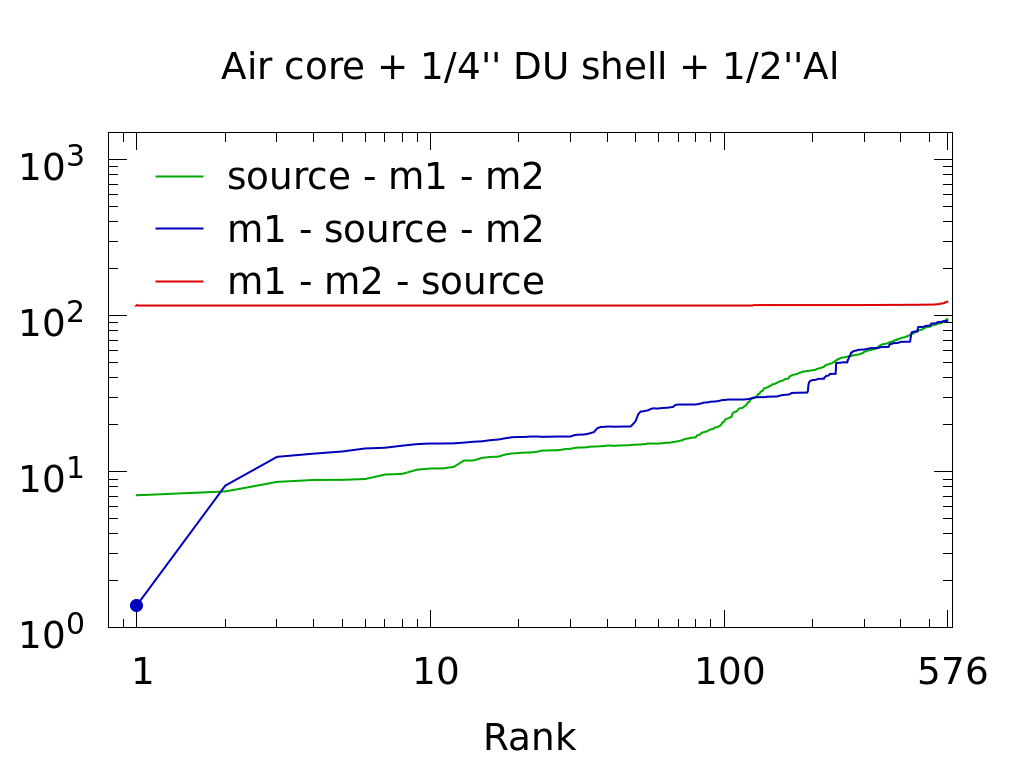} &
      \includegraphics[width=0.33\textwidth]{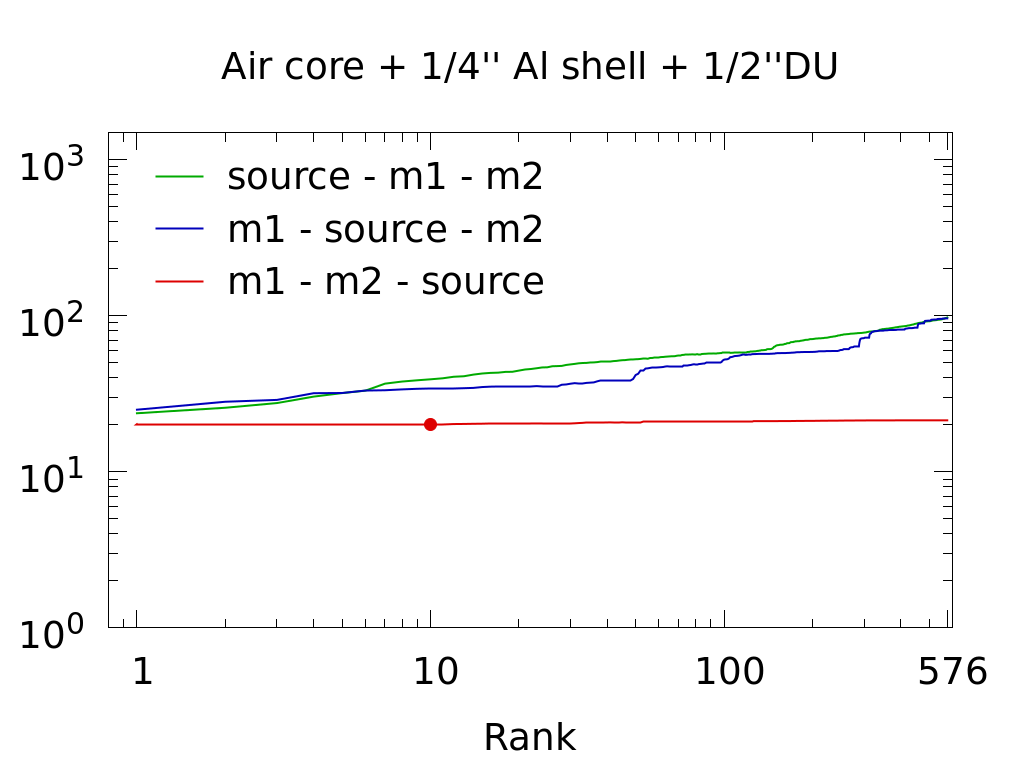}
    \end{tabular}
    \caption{Loss function as a function of rank for three cases with three permutations of materials, including the source shell.
    Geometry is kept identical in all three cases: spherical core + 1/4'' shell + 1/2'' shell.
    In the left, middle, and right panels, the source material is located in the core, middle shell, and outer shell, respectively.
    Dots mark location of the ground truth with the respective color.
    In all cases, ground truth shows the lowest loss function for the correct sequence of materials.
    } 
    \label{fig:identifying-source}
\end{figure*}

Finally, we turn to the problem of identifying which shell is the radioactive source.
In the examples above, we assumed that source was located in the innermost shell. 
The analytic expressions derived in~\ref{app:analytic-deriv} are valid for general case, in which the source is not necessarily the innermost shell.
In such case, the location of the source can be deduced by the brute-force approach, where we simply try an assumption that the source is a specific shell and see which of the trials leads to the minimal loss function. 
The resulting trends of the loss function sorted by rank reveal the correct source location, see Figure~\ref{fig:identifying-source}.

\section{Conclusion}
\label{sec:conclusion}
A novel brute-force statistical method has been developed to perform
material identification for an arbitrary number of materials using a high spectral resolution detector, such as HPGe.
The key ingredient in this method is the use of semianalytic expressions to evaluate the uncollided flux from a system of nested shielding / source shells in spherical or cylindrical geometries (see~\ref{app:analytic-deriv}).
The method is applicable to cases in which the materials to be identified are within a finite set of possible candidates (as opposed to a continuum of materials, characterized by e.g. an effective atomic number $Z$).
The method has been tested on a suite of previously published experimental data with shielded depleted uranium shells as well as a Beryllium Reflected Plutonium (BeRP) ball (see Table~\ref{tab:experiments}).
To diagnose potential sources of error, the method has also been applied to simulated spectra and uncollided flux computed for the same types of objects.

The results show the method to be very successful when a single or a pair of materials need to be identified.
For the case of three and more materials, the method gives a list of top possible candidates, and other means may needed to further refine the solution space.
Specifically, when using peak areas for line counts in simulated spectra, ground truth combinations score in the top 3\%, 1.5\%, and 0.2\% of all the combinations for three, four, and five unknown materials, respectively.
For the experimental data with DU shells, the ground truth is within the top 0.1\%.
The same holds true for BeRP ball with a single shielding shell when it is split into 3 or 4 subshells.
However, material identification using the BeRP ball and polyethylene or mock high-explosive shells performs worse, giving ground truth scores only within the top 20\%.

Although the method relies on prior knowledge of the geometry of the system, which can be obtained as a preliminary step using e.g. radiography, it can also be applied to probe this geometry to a certain extent.
For example, if the boundaries between the shielding shells are not known, the shielding can be split into a small number of thin shells, and material ID can be performed on each of the subshells individually.
Thicknesses of contiguous spans of the same material then give approximate thickness of the shielding made of that material.
We foresee this approach, however, to have some limitations in terms of how thin these subshells are allowed to be.
This is not only because the computational cost increases as a factorial of the number of unknown shells, but also thinner shells with small areal mass produce insufficient imprint on the line spectrum, creating more confusion.

The method can be potentially improved by performing more advanced continuum subtraction, as well as considering correlated attenuation for different lines in the statistical model.
The method can also be extended to general 3D geometries, which will be the topic of future studies.

\section{Acknowledgements}

The authors would like to acknowledge the funding for this research provided by the National Nuclear Security Administration Office NA-22 and the resources provided by the Los Alamos National Laboratory (LANL), USA.
In particular, we used resources provided by the LANL Institutional Computing Program. 
LANL is operated by Triad National Security, LLC, for the National Nuclear Security Administration of the US DOE (Contract No. 89233218CNA000001). 
This work is authorized for unlimited release under LA-UR-23-29990.


\appendix
\section{Derivation of analytic expressions for uncollided flux}
\label{app:analytic-deriv}

\begin{figure}[!htbp]
    \centering
    \includegraphics[width=\columnwidth]{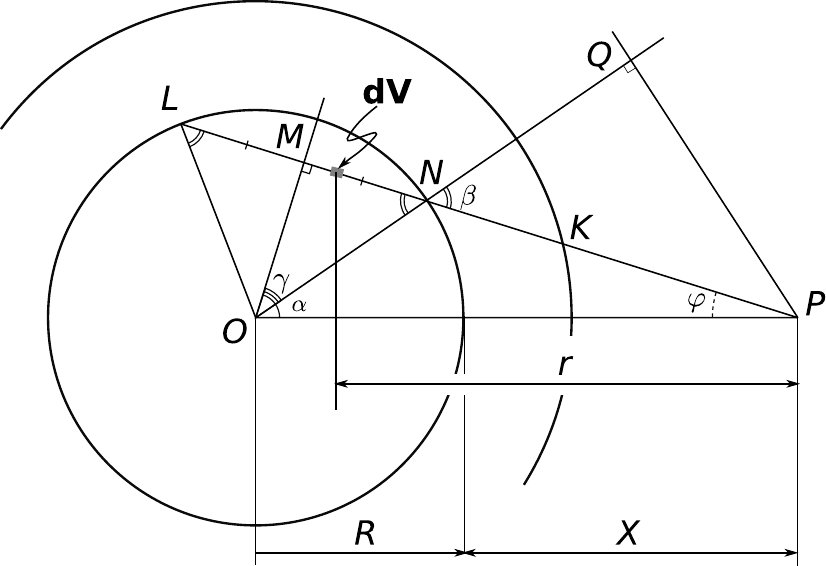}
    \caption{Sketch illustrating integration angles.}
    \label{fig:integration-sketch}
\end{figure}

For specific case of the onion-shaped spherical and cylindrical configurations with arbitrary number of concentric shells, general expressions can be derived.
We will assume that each shell is composed of a single uniform material, and one of the shells is a radioactive source of gamma-emission. 
In our setting, the source shell need not be the innermost one.
We ignore scattering and assume that every shell provides a unique energy-dependent profile of attenuation $\mu_k(E)$, dictated by the constituent material.
The gamma-emission is then observed from a distance $X_0$ to the edge of the radiating shell at a point $P$ (see Fig.~\ref{fig:integration-sketch}).
 
For clarity, we start with a single spherical source of radius $R_0$ and attenuation coefficient $\mu_0$, surrounded by a shell of shielding material of radius $R_1$ and attenuation coefficient $\mu_1$.
Previous study~\cite{tsuruo1965} provides the following starting expression for the emergent flux $F$ at a point $P$, located at distance $X$ from the surface of the spherical source:
\begin{align}
    F &= \iiint \frac{\cos{\varphi}\ dV}{4\pi\rho^2} 
         S_v e^{-\mu_0(r-r_1)\sec{\varphi} - \mu_1 d_1} 
         \label{eq:emflux}.
\end{align}
Here, $\rho$ is the distance from an integration volume $dV$ to the detector $P$, $r$ ($r_1$) is the projection of $\rho$ (NP) onto the horizontal line $OP$, such that $r = \rho \cos\varphi$ ($r_1 = |NP|\cos\varphi$), $d_1=|NK|$, and $S_v$ is the gamma-ray source intensity.
An additional factor of $\cos\varphi$ here accounts for an angle of incidence (missing in the starting equation 3 in Tsuruo 1965~\cite{tsuruo1965}). 
Writing the volume element in $(r,\varphi)$-coordinates and integrating out the $r$ variable, we get
\begin{align}
    F  &= \int_0^{\varphi_0}  
          \int_{r_1(\varphi)}^{r_2(\varphi)} 
          \frac{\cos{\varphi}}{4\pi\rho^2}
          \left(2\pi\rho^2\sin{\varphi}\,d\varphi\,\frac{dr}{\cos{\varphi}}\right)
         \nonumber\\
         &\times S_v e^{-\mu_0(r-r_1)\sec{\varphi} - \mu_1 d_1} =
         \nonumber\\
      &= \frac{S_v}{2}\int_0^{\varphi_0} d\varphi \sin{\varphi}
         \int_{r_1}^{r_2} dr
         e^{-\mu_0(r-r_1)\sec{\varphi} - \mu_1 d_1} =
         \nonumber\\
      &= \frac{S_v}{2\mu_0} \int_0^{\varphi_0} \cos{\varphi}\,d(\cos{\varphi})
         e^{-\mu_1 d_1}
         \left(1 - e^{-\mu_0\,d_0}\right).
         \label{eq:flux-int-varphi}
\end{align}
In this expression, $r_1(\varphi)$ and $r_2(\varphi)$ are the $\varphi$-dependent integration limits of $r$, and $d_0$ is the length of the segment LN.

It is convenient to change to an integration angle $\gamma$ instead of $\varphi$ (see Fig.~\ref{fig:integration-sketch}):
\begin{align}
    \gamma = \frac{\pi}{2} - \beta = \frac{\pi}{2} - \alpha - \varphi,
\end{align}
since the length $d_0$ is much easier to express in terms of the angle $\gamma$:
\begin{align}
    d_0 = |LN| = 2 R_0 \sin{\gamma}.
\end{align}
The angle $\varphi$ can be related to $\gamma$ as follows:
\begin{align}
    &\sin{\varphi} = \frac{R_0\cos{\gamma}}{R_0 + X_0} = \frac{\cos\gamma}{1 + \xi}, \\
    &\cos\varphi\, d\cos\varphi = \frac{\sin\gamma\, d\sin\gamma}{(1+\xi)^2}.
\end{align}
where $\xi \equiv X_0/R_0$. 
Expression for the flux becomes 
\begin{align}
    F = \frac{S_v}{2\mu_0(1+\xi)^2}\int_0^{\pi/2} 
        &e^{-\mu_1 d_1(\gamma)}
        \left(1 - e^{-2\mu_0 R_0\sin\gamma}\right)\nonumber\\
        &\times\sin\gamma\, d\sin\gamma,
    \label{eq:flux-sphere}
\end{align}
where the limits of integration change from $\{0,\varphi_0\}$ in (\ref{eq:flux-int-varphi}) to $\{0,\pi/2\}$.
If we ignore the term with $\mu_1 d_1$ (a case with no shielding), then after the change of integration variable to $x:=\sin\gamma$ the flux is
\begin{align}
    F = \frac{S_v}{2\mu_0(1+\xi)^2}\int_0^1 
             {x\, dx\left(1 - e^{-2\mu_0 R_0 x}\right)}.
\end{align}
Integration can be done in closed form. The resulting formula is equivalent to Equation~(19) in Atkinson~\cite{atkinson2006}. 
\begin{align}
    F &= \frac{S_v (1 + 2x_0(1+x_0)e^{-1/x_0} - 2x_0^2)}{4\mu_0(1+\xi)^2},
    \label{eq:flux-sphere2}
\end{align}
where $x_0:=(2\mu_0 R_0)^{-1}$. 

\subsection*{Multiple shells}

The central sphere with radioactive source has material extinction coefficient $\mu_0$ and radius $R_0$. 
Consider a sequence of nested spherical shielding shells, with extinction coefficients $\mu_1,\dots,\mu_n$ and outer radii $R_1,\dots,R_n$.
Each outer shell adds a multiplicative extinction factor $e^{-\mu_k d_k}$, where $d_k$ is the length traveled in the $k$-th shell along the ray.

In particular, in Figure~\ref{fig:integration-sketch}, $d_1 = |NK|$.
In analogy with the angle $\gamma$, we can introduce angles $\gamma_1 := \angle MOK$, $\gamma_2,\dots,\gamma_n$, between the ray $OM$ and the ray originating from $O$ and passing through the intersection point with the $j$-th shell. 
These angles are related by the chain of identities:
\begin{align}
    R_0\cos\gamma = R_1\cos\gamma_1 = \dots = R_n\cos\gamma_n.
\end{align}

The lengths $d_j$ can be computed as:
\begin{align}
    d_0 &= 2R_0\sin\gamma, \nonumber \\
    d_1 &= R_1\sin\gamma_1 - R_0\sin\gamma, \nonumber \\
    d_2 &= R_2\sin\gamma_2 - R_1\sin\gamma_1, \nonumber \\
    \dots \nonumber \\
    d_n &= R_n\sin\gamma_n - R_{n-1}\sin\gamma_{n-1}, \nonumber
\end{align}
which can all be expressed in terms of $x:=\sin\gamma$:
\begin{align}
    d_0 &= 2R_0 x, 
    \label{eq:dees} \\
    d_1 &= R_1\sqrt{1 - \zeta_1^2(1-x^2)} - R_0 x, \nonumber \\
    d_2 &= R_2\sqrt{1 - \zeta_2^2(1-x^2)} - R_1\sqrt{1 - \zeta_1^2(1-x^2)}, \nonumber \\
    \dots \nonumber \\
    d_n &= R_n\sqrt{1 - \zeta_n^2(1-x^2)} - R_{n-1}\sqrt{1 - \zeta_{n-1}^2(1-x^2)},
    \nonumber
\end{align}
where $\zeta_j:=R_0/R_j$ for $j = 1,\dots,n$.

Correspondingly, expression (\ref{eq:flux-sphere}) generalizes to:
\begin{align}
    F = \frac{S_v}{2\mu_0(1+\xi)^2}
     \int_0^1 &x\, dx\left(1 - e^{-2\mu_0 R_0 x}\right)\nonumber\\
     &\times\exp{\left\{-\sum_{j=1}^{n}\mu_j d_j(x)\right\}}.
    \label{eq:fluxfinal}
\end{align}

\subsection*{Radioactive shell not the innermost one}

Calculations above can be extended to include the case where the shell that is the source of radiation is not the innermost one.
In this case, the flux can be split into two contributions: one generated in the hemisphere closer to the detector and only passing through outer shells, and one that is produced in the farther hemispheric region, that passes through all the inner shells as well.
Let the $m$-th shell be the one with radioactive source, and  $m\ge1$.
The total flux at the detector location becomes:
\begin{align}
    F &= \frac{S_v}{2\mu_m(1+\xi_m)^2}
     \int_0^1 y\, dy\left(1 - e^{-\mu_m d_m(y)}\right)\nonumber\\  
     &\times\exp{\left\{-\sum_{j=m+1}^{n}\mu_j d_j(y)\right\}}
     \nonumber \\
     &\times
     \left[1 + \exp{\left\{-\mu_m d_m(y) - 2\sum_{j=1}^{m-1}\mu_j d_j(y)\right\}}
     \right]
    \label{eq:flux-shell-not-innermost}
\end{align}
where $y:=\sin\gamma_m$, $\xi_m:=X_m/R_m$, and
\begin{align}
    d_m(y) = 
    \begin{cases}
      R_m y & \text{if } y \le \sin\gamma_m, \\
      R_m y - R_{m-1}\sqrt{1 - \zeta_{m+1}^2(1 - y^2)} & \text{if } y > \sin\gamma_m,
    \end{cases}
\end{align}
with $\zeta_m := R_{m-1}/R_m$.
In this expression for the flux, the two terms in the square brackets, the unit and the exponent, correspond to the emission from the near and far side of the shell, respectively.

\subsection*{Cylindrical geometry}

In case of cylindrical geometry, we will ignore the lateral distortion coming from the finite height of the cylinder $H$, and simply consider the same model in 2D. 
This approximation is particularly applicable when the cylinder is flat, i.e. $H\ll R$.
For a cylindrical source with attenuation $\mu_0$, surrounded by a shell with attenuation $\mu_1$, an expression for the flux similar to Equation~(\ref{eq:emflux}) is:
\begin{align}
    F &= \iint \frac{\cos\varphi H dA}{4\pi\rho^2} 
         S_v e^{-\mu_0(\rho-\rho_1) - \mu_1 d_1} =
         \\
      &= \frac{S_v\,H}{4\pi}\int_0^{\varphi_0}  
          e^{-\mu_1 d_1 + \mu_0 \rho_1}
          \cos{\varphi}\,d\varphi
          \int_{\rho_1(\varphi)}^{\rho_2(\varphi)} 
          \frac{e^{-\mu_0\rho}d\rho}{\rho},
\end{align}
where $\varphi_0 = \arcsin{(\xi+1)^{-1}}$, and the limits of integration over $\rho$ are
\begin{align}
    \rho_{1,2} &= R_0\left((\xi + 1)\cos\varphi
                          \mp \sqrt{(\xi + 1)^2\cos^2\varphi - \xi(\xi+2)}\right).
\end{align}
The integral over $\rho$ is a familiar special function known as the \emph{exponential integral}, $Ei(z) := \int_z^\infty e^{t}dt/t$. We can use it to formulate the final expression for the flux as
\begin{align}
    F  = \frac{S_v\,H}{4\pi\mu_0}\int_0^{\varphi_0}  
          &e^{-\mu_1 d_1 + \mu_0 \rho_1}\nonumber\\
          &\times\left(Ei(-\mu_0 \rho_2) - Ei(-\mu_0 \rho_1)\right)
          \cos{\varphi}\,d\varphi.
    \label{eq:flux-flat-cyl}
\end{align}

It can also be generalized to the case when there are multiple shells, and the source shell is at the $m$-th position, similarly to the expression (\ref{eq:flux-shell-not-innermost}) for spheres:
\begin{align*}
    F &= \frac{S_v\,H}{4\pi\mu_m}
     \int_{\varphi_{m-1}}^{\varphi_m} 
     d\varphi\,\cos{\varphi}\,
     e^{\mu_m \rho^{(m)}_1}
     \left(Ei(-\mu_m \rho^{(m)}_2) - Ei(-\mu_m \rho^{(m)}_1)\right)
     \exp{\left\{-\sum_{j=m+1}^{n}\mu_j d_j(\varphi)\right\}} +
     \nonumber \\
     &+ \frac{S_v\,H}{4\pi\mu_m}
     \int_{0}^{\varphi_{m-1}}
     d\varphi\,\cos{\varphi}\times
     \nonumber \\
     &\times \left[ e^{\mu_m \rho^{(m)}_1}
     \left(Ei(-\mu_m \rho^{(m-1)}_1) - Ei(-\mu_m \rho^{(m)}_1)\right)
     \exp{\left\{-\sum_{j=m+1}^{n}\mu_j d_j(\varphi)\right\}} + \right.
     \nonumber \\
     &+ \left.e^{\mu_m \rho^{(m-1)}_2}
     \left(Ei(-\mu_m \rho^{(m)}_2) - Ei(-\mu_m \rho^{(m-1)}_2)\right)
     \exp{\left\{-\sum_{j=m+1}^{n}\mu_j d_j(\varphi) - 2\sum_{j=1}^{m-1}\mu_j d_j(y)\right\}}
     \right]
\end{align*}
where $\xi_m:=X_m/R_m$, $\varphi_m = \arcsin{(\xi_m+1)^{-1}}$, and
\begin{align}
    &\rho^{(k)}_{1,2} = R_k\left((\xi_k + 1)\cos\varphi
                    \mp \sqrt{(\xi_k + 1)^2\cos^2\varphi - \xi_k(\xi_k + 2)}\right), 
    \nonumber \\
    &d_k(\varphi) = 
    \begin{cases}
    \rho^{(k)}_2 - \rho^{(k-1)}_2 \equiv \rho^{(k-1)}_1 - \rho^{(k)}_1
    & {\rm if}\; \varphi < \varphi_{k-1}, 
    \\
    \frac12(\rho^{(k)}_2 - \rho^{(k)}_1)
    & {\rm if}\; \varphi_{k-1}\le\varphi < \varphi_{k}, 
    \\
    0 & {\rm if}\; \varphi_{k} \le \varphi. 
    \end{cases}
\end{align}

This expression is significantly more complex than for the spherical case. 
Nevertheless, it only involves a single integration, which can be computed numerically with high speed and efficiency.
Notice that, unlike spherical case, the flux in (\ref{eq:flux-shell-not-innermost}) does not show simple inverse-square scaling $(1+\xi)^{-2}$ with distance to detector $\propto\xi$ (although it can be shown that it asymptotically approaches such scaling for $\xi\to\infty$).

\end{document}